\documentclass{elsart}
\usepackage{natbib}
\usepackage{amsmath}
\usepackage{amssymb}
\usepackage{amsfonts}
\usepackage{dsfont}
\usepackage{epsfig}
\usepackage{graphicx}
\usepackage{relsize}
\usepackage{txfonts}
\usepackage{wasysym}
\usepackage{tocloft}

\newcommand{\dsty}[1]{\displaystyle{#1}}

\graphicspath{{./FIGURES/}}
\begin{document}
\begin{frontmatter}
\title{On the consistency of MPS}
\author[label1]{Antonio Souto-Iglesias\corauthref{cor}},
\corauth[cor]{Corresponding author.} \ead{antonio.souto@upm.es}
\author[label1]{Fabricio Maci\`{a}},
\author[label1]{Leo M. Gonz\'{a}lez},
\author[label1]{Jos\'{e} L. Cerc\'{o}s-Pita},
\address[label1]{Naval Architecture Department (ETSIN),\\
Technical University of Madrid (UPM), 28040 Madrid, Spain.}
\begin{abstract}
The consistency of Moving Particle Semi-implicit (MPS) method in
reproducing the gradient, divergence and Laplacian
differential operators is discussed in the present paper.
Its relation to the Smoothed Particle Hydrodynamics (SPH) method is
rigorously established. The application of the MPS method to solve
the Navier-Stokes equations using
a fractional step approach is treated, unveiling
inconsistency problems when solving
the Poisson equation for the pressure.
A new corrected MPS method incorporating boundary terms is proposed.
Applications to one dimensional boundary value
Dirichlet and mixed Neumann-Dirichlet problems and to
two-dimensional free-surface flows are presented.
\end{abstract}
\end{frontmatter} 
\section{Introduction}
The ``Moving Particle Semi-implicit'' method (MPS) is a numerical technique first introduced by Koshizuka and Oka in the nineties \citep{koshizuza1995,koshizuka1996} to solve incompressible Navier-Stokes equations. It relies on a meshless discretization of the continuum
to approximate differential operators. In order for the obtained flow field to exhibit the properties of an incompressible flow, a fractional time step approach is used. This approach requires solving a Poisson equation for the pressure field at every time step, which consequently demands a numerical approximation of the Laplacian operator.

The MPS scheme is mainly used to solve violent free-surface flows, exploiting the fact that no mesh is needed for the computations. This can be seen in \citep{koshizuka1998} where the technique was used to simulate breaking waves on a sloping beach and in
\citep{Khayyer_gotoh_ce09,Khayyer_gotoh_ijope09} where impact pressures in sloshing flows are accurately computed.
\citet{youngyoon1999b} extended MPS method to simulate multiphase flows and
\citet{youngyoon1999} improved the MPS formulation in regards to the advection scheme by incorporating an arbitrary Lagrangian-Eulerian model.

With the increasing computational power in the early 2000s, Naito and Sueyoshi \citep{naito2001,sueyoshi2003} extended the method to 3D.  They utilized it to simulate complex marine engineering problems, such as the sinking of a vessel after the flooding of the cargo holds and green water events.
\citet{Tsukamoto_cheng_2011} adapted the MPS method to model complex fluid structure interaction problems.
Different corrections aimed at improving
the accuracy of the method have recently been documented.
\citet{Tanaka_etal_jcp2010_mps} introduced some corrections in the fractional
step pressure solver in order to mitigate pressure oscillations.
\citet{Khayyer_gotoh_jcp_2012_mps} enhanced the capabilities of the method 
in order to better approximate
tensile states in fluids and developed a
higher order approximation to the differential operators
aiming at obtaining more accurate pressure 
fields \citet{Khayyer2010_aor,Khayyer2012_aor}; the improvement
is demonstrated with results for sloshing flows both in two and three dimensional cases.

Regardless all that body of work neither the derivation of the MPS numerical method from its first principles,
nor its consistency in regards to how accurately it reproduces the continuum
as the numerical resolution increases, have been discussed at length in the literature.
As CFD practitioners we believe this issue is paramount in order to guarantee the Engineering applicability of the numerical technique.

The present paper addresses this matter by first analyzing the consistency of the MPS approximations to the gradient,
divergence and Laplacian operators and deriving continuous integral forms of these approximations.
Second, the connections with Smoothed Particle Hydrodynamics (SPH),
see e.g. \citep{monaghan_arfm_2012},
are rigorously presented by relating, through the integral form of the operators,
the MPS weighting function with the SPH kernel.
Third, the integral form of the operators is applied to the solution of the Poisson
equation used in MPS for the pressure computation through the fractional step methodology.
Some problematic issues regarding the implementation
of Dirichlet and Neumann boundary conditions are discussed and
some corrections to the operators proposed.
Finally, applications to one-dimensional problems and two-dimensional
free-surface flows are presented. 
\section{Governing equations}
\label{s:governing_equations}
Newtonian incompressible flows are the ones treated with the MPS method.
The incompressible Navier-Stokes equations
in Lagrangian formalism are hence taken as the field equations:
\begin{align}
\label{Navier_Stokes_inc1}
\frac{D \mathbf{r}}{D t} \, = &\, \mathbf{u}, \\
\label{Navier_Stokes_inc2}
\dsty{\nabla\cdot}\mathbf{u}  \, =& \, 0, \\
\dsty{\frac{D \mathbf{u}}{D t}} \, = &\, \mathbf{g} \, + \, \frac{\nabla \cdot \varmathbb{T}}{\rho}.
\label{Navier_Stokes_inc}
\end{align}
where $\rho$ stands for the fluid density  and $\mathbf{g}$ is a generic
external volumetric force field. The flow velocity $\mathbf{u}$ is defined as the material
derivative of a fluid particle with position $\mathbf{r}$.
$\varmathbb{T}$ denotes the stress tensor of a Newtonian incompressible fluid:
\begin{equation}
\label{Tstress}
\varmathbb{T}  \, =  - P \,\boldsymbol{\mathit{I}}\, + \, 2 \, \mu \, \varmathbb{D} \,,
\end{equation}
in which $P$ is the pressure, $\varmathbb{D}$ is the rate of deformation tensor
$(\varmathbb{D} = ( \nabla \mathbf{u} + \nabla \mathbf{u}^{T} )/2)$
and $\mu$ is the dynamic viscosity.
With this notation, the divergence of the stress tensor $\varmathbb{T}$
is computed as:
\begin{equation}
\label{varmathbbT}
\nabla \cdot \varmathbb{T} \, = \, - \, \nabla P \,  + \, \mu \nabla^2 \mathbf{u} \,.
\end{equation}
The MPS method is based on a Helmholtz-Hodge decomposition of an intermediate velocity field initially
devised by Chorin \citep{chorin_fractional_step_poisson_1968} in the late sixties.
First, an intermediate velocity field $\mathbf{u^*}$ is explicitly computed using
the momentum equation but ignoring the pressure term. Second, the zero divergence condition
is imposed on the velocity field at
the next time step, thus obtaining the Poisson equation for the pressure:
\begin{equation}
\label{eq:poissonMPS}
\nabla^2 P=\frac{\rho}{\Delta t}\left(\nabla\cdot\mathbf{u^*}\right),
\end{equation}
in which $\Delta t$ is the time step. Once the pressure is found, pressure gradients
and particle positions are modified.
A comprehensive flow-chart of the whole
approach can be found in e.g. \citep{youngyoon1999}.

\section{MPS discretization}
\label{s:diff_operators}
\subsection{Density and number density}
\label{s:weightf}
The set of equations (\ref{Navier_Stokes_inc1}-\ref{Navier_Stokes_inc})
has to be discretized in order to implement the fractional step algorithm.
The fluid domain is discretized in a set of particles and the 
differential operators at an individual 
particle $i$ are evaluated using the value of the different fields at the
neighboring particles. The MPS method weights these neighboring particle properties
by using a compactly supported function $w$ whose
argument is the distance between particles
$\left|\mathbf{x}_i-\mathbf{x}_j\right|$ normalized
with the cut-off radius $r_e$ and which is non-negative.
The weighting function is singular at the
origin in some formulations \citep{koshizuka1996,Tsukamoto_cheng_2011,Khayyer_gotoh_jcp_2012_mps}
and regular in others \citep{youngyoon1999,youngyoon1999b}.
The notation of reference \citep{youngyoon1999} is used to present the formalism.


A particle number density at a particle with 
coordinates $\mathbf{x}_i$ is defined as:
\begin{equation}
\label{eq:ni}
\langle n \rangle_i :=
\sum_{j\in J_i}
w
\left(
        \frac
        {\left|\mathbf{x}_i-\mathbf{x}_j\right|}
        {r_e}
\right),
\end{equation}
where $J_i$ consists in all neighboring particles' indexes and 
it may or may not include $i$ itself \citep{youngyoon1999b,youngyoon1999}.

In the MPS literature, only the dependance of $w$ on the distance $\left|\mathbf{x}_i-\mathbf{x}_j\right|$ is made explicit in the
notation. However, the definition of the weighting function incorporates the cut-off radius $r_e$ in terms of the ratio $\left|\mathbf{x}\right|/r_e$.
We have chosen to make explicit the dependance of the weighting function on $r_e$; the benefits of this choice, which is non-standard in the
MPS literature, will be clear in the computations that follow.

The approximation to the density field for a particle $i$ is obtained from
$\langle n \rangle_i$ as \citep{youngyoon1999}:
\begin{equation}
\label{eq:mpsdensity}
\langle \rho \rangle_i=
\frac
{
    m \langle n \rangle_i
}
{
\int_{\mathbb{R}^d}
w
\left(
        \frac
        {\left|\mathbf{x}\right|}
        {r_e}
\right)
\,d\mathbf{x}
},
\end{equation}
where $m$ is the mass of each particle and $d$ is the number of dimensions.

Since we are solving incompressible flows, the fluid has its
constant reference density $\rho_0$ as a fundamental
physical magnitude, for which a compatible reference number density $n_0$
can be defined from equation (\ref{eq:mpsdensity}) as:
\begin{equation}
\label{eq:n0}
n_0 =
\frac
{
    \rho_0
}
{
m
}
\,
\int_{\mathbb{R}^d}
w
\left(
        \frac
        {\left|\mathbf{x}\right|}
        {r_e}
\right)
\,d\mathbf{x}.
\end{equation}
The mass of a particle $m$ can be obtained as the total mass of the fluid $M$
divided by the number of particles $N$. Moreover, the volume integral of the kernel
in equation (\ref{eq:n0}) equals, by defining $\mathbf{q}=\mathbf{x}/r_e$:
\begin{equation}
\int_{\mathbb{R}^d}
w
\left(
        \frac
        {\left|\mathbf{x}\right|}
        {r_e}
\right)
\,d\mathbf{x}
=
\left(r_e\right)^d
\int_{\mathbb{R}^d}
w
\left(
        \left|\mathbf{q}\right|
\right)
\,d\mathbf{q}.
\end{equation}
Let us define $A_0$ as
\begin{equation}
\label{eq:A_0}
A_0=
\int_{\mathbb{R}^d}
w
\left(
        \left|\mathbf{q}\right|
\right)
\,d\mathbf{q}.
\end{equation}
This value only involves the shape of the kernel
and does not depend on the number of particles
nor does it depend on  the specific value of $r_e$.
Using this notation, equation (\ref{eq:n0}) becomes:
\begin{equation}
\label{eq:n0a}
n_0 =
\frac
{
    \rho_0
}
{
(M/N)
}
\,
\left(r_e\right)^d\, A_0.
\end{equation}
The ratio $M/\rho$ of the total mass over the density is the total volume $V$. Therefore,
\begin{equation}
\label{eq:n02}
n_0 =
\frac
{
    1
}
{
(V/N)
}
\,
\left(r_e\right)^d\, A_0.
\end{equation}
The total volume over the number of particles, $V/N$, is the volume $\upsilon$ associated to each particle.
Let us call $\Delta x$ the typical particle spacing. It can be assumed that $\upsilon=\left(\Delta x\right)^d$
and with this in mind the value $n_0$ satisfies
\begin{equation}
\label{eq:1overn0}
\frac
{
    1
}
{
    n_0
}
=
\frac{1}{A_0}
\left(
\frac{\Delta x}{r_e}
\right)
^
d.
\end{equation}
This expression will be useful in the computations that follow.

The ratio $r_e/\Delta x$ is an indicator of the number of particles inside the weighting function support (neighbors).

Identity (\ref{eq:1overn0}) is important because it allows for the introduction
of the typical particle distance $\Delta x$ in the formalism.
The integral version of the MPS differential
operators is obtained by taking the limit as $\Delta x\rightarrow 0$.
%

\subsection{Gradient and divergence}
\label{ss:mpsgrad}
The gradient in the MPS method for a scalar
function $\phi$ at a particle with co-ordinates $\mathbf{x}_i$
is approximated (see e.g. \citep{youngyoon1999}) by the following sum
\begin{equation}
\label{eq:mpsgraddiscr}
\langle \nabla \phi \rangle_i
=
\frac{d}{n_0}
\sum_{j\neq i}
\left[
    \frac
    {
        \phi_j - \phi_i
    }
    {
        {\left|\mathbf{x}_j-\mathbf{x}_i\right|^2}
    }
    \left(\mathbf{x}_j-\mathbf{x}_i\right)
    w
    \left(
            \frac
            {\left|\mathbf{x}_j-\mathbf{x}_i\right|}
            {r_e}
    \right)
\right].
\end{equation}
This sum can be interpreted as a discrete approximation of an integral.
Taking into account the value of $n_0$ given by equation (\ref{eq:1overn0})
a continuum analogue of the MPS discrete gradient is defined by:
\begin{equation}
\label{eq:partial_xk_phi}
\left\langle \partial_{x_{k}}\phi\right\rangle \left(  \mathbf{x}\right)
\approx
\frac{d}{\left(r_e\right)^d A_0}
\int_{\mathbb{R}^d}
    \frac
    {
        \phi(\mathbf{x^{\prime}}) - \phi(\mathbf{x})
    }
    {
        {\left|\mathbf{x^{\prime}}-\mathbf{x}\right|^2}
    }
    \left(x^{\prime}_k-x_k\right)
    w
    \left(
            \frac
            {\left|\mathbf{x}-\mathbf{x^{\prime}}\right|}
            {r_e}
    \right)
    d\mathbf{x^{\prime}}.
\end{equation}
The question arises whether this expression is a consistent approximation to the gradient of a function.
It is possible to compare it with the SPH expression of the gradient
(see section \ref{ss:appendix:grad} in the appendix),
whose consistency has been established in the literature (see appendix for details).
Let us rewrite (\ref{eq:partial_xk_phi}) as:
\begin{equation}
\label{eq:mpsgrad}
\left\langle \partial_{x_{k}}\phi\right\rangle \left(  \mathbf{x}\right)
\approx
-
\frac{d}{\left(r_e\right)^d A_0}
\int_{\mathbb{R}^d}
    \frac
    {
        \phi(\mathbf{x^{\prime}}) - \phi(\mathbf{x})
    }
    {
        {\left|\mathbf{x}-\mathbf{x^{\prime}}\right|}
    }
    \left(x_k-x^{\prime}_k\right)
    \frac
    {
        1
    }
    {
        {\left|\mathbf{x}-\mathbf{x^{\prime}}\right|}
    }
    w
    \left(
            \frac
            {\left|\mathbf{x}-\mathbf{x^{\prime}}\right|}
            {r_e}
    \right)
    d\mathbf{x^{\prime}}.
\end{equation}
If expressions (\ref{eq:mpsgrad}) and (\ref{eq:contsphgrad})
in the appendix were to coincide for every function $\phi$,
necessarily, the following relation between the SPH kernel and the given
MPS weighting function must hold:
\begin{equation}
\label{eq:hreWw}
    \frac
    {
        1
    }
    {
        h^{d+1}
    }
    \tilde{W}^{\prime}
    \left(
            \frac
            {\left|\mathbf{x}\right|}
            {h}
    \right)
=
-
\frac{d}{\left(r_e\right)^d A_0}
    \frac
    {
        1
    }
    {
        {\left|\mathbf{x}\right|}
    }
    w
    \left(
            \frac
            {\left|\mathbf{x}\right|}
            {r_e}
    \right).
\end{equation}
Assuming that the MPS weighting function and the SPH kernel
characteristic lengths are equal ($h \equiv r_e$), we have
$$\mathbf{q}= \frac{\mathbf{x}}{r_e}= \frac{\mathbf{x}}{h}.$$
SPH practitioners usually consider kernels with a $2h$ support radius whilst the radius of the MPS
weighting function is $r_e$. Nonetheless, for the sake of simplicity, we will assume that both are equal.

Set $q:=\left|\mathbf{q}\right|$; equation (\ref{eq:hreWw}) can then be rewritten as:
\begin{equation}
\label{eq:Ww}
    \tilde{W}^{\prime}(q)
=
-
\frac{d}{A_0}
    \frac{1}{q}w(q).
\end{equation}
This expression can be used
to obtain the function $\tilde{W}$ involved
in the definition of the SPH kernel from the MPS weighting function $w$.
However, in order for equation (\ref{eq:Ww}) to give rise to a well defined
SPH kernel, we must first check that:
\begin{equation}
\label{eq:kernelint1}
\int_{\mathbb{R}^{d}}\tilde{W}\left(  \left\vert \mathbf{q}\right\vert
\right)  d\mathbf{q}=1.
\end{equation}

This is indeed the case since combining equations (\ref{eq:A_0}), (\ref{eq:Ww}) and (\ref{eq:wgradw_d}) from
the appendix, we deduce that
\begin{displaymath}
\int_{\mathbb{R}^d}
    \tilde{W}(\left|\mathbf{q}\right|) d\mathbf{q}
=
-
\frac{1}{d}
\int_{\mathbb{R}^d}
\left|\mathbf{q}\right|\tilde{W}^{\prime}(\left|\mathbf{q}\right|)d\mathbf{q}
=
\frac{1}{A_0}
\int_{\mathbb{R}^d}
w(\left|\mathbf{q}\right|)d\mathbf{q}
=
1.
\end{displaymath}
%
%
Let us now focus on the divergence operator, with $\mathbf{u}$ being  a generic vector field.
The divergence of $\mathbf{u}$ at a particle with co-ordinates
$\mathbf{x}_i$ is computed through the MPS method (see e.g. \citep{youngyoon1999}) with the following sum:
\begin{equation}
\langle \nabla \cdot \mathbf{u} \rangle
=
\frac{d}{n_0}
\sum_{j\neq i}
\left[
    \frac
    {
        \left(
            \mathbf{u}_j - \mathbf{u}_i
        \right)
        \cdot
        \left(
            \mathbf{x}_j - \mathbf{x}_i
        \right)
    }
    {
        {\left|\mathbf{x}_j-\mathbf{x}_i\right|^2}
    }
    w
    \left(
            \frac
            {\left|\mathbf{x}_j-\mathbf{x}_i\right|}
            {r_e}
    \right)
\right].
\end{equation}
Substituting $n_0$ with its value from equation (\ref{eq:1overn0}) and interpreting 
the sum as a discrete approximation of an integral, we get a continuum analogue of the MPS discrete divergence:
\begin{equation}
\label{eq:mpsdivvv}
\left\langle \nabla \cdot \mathbf{u}  \right\rangle \left(  \mathbf{x}\right)
\approx
\frac{d}{\left(r_e\right)^d A_0}
\int_{\mathbb{R}^d}
    \frac
    {
        \left(
            \mathbf{u^\prime} - \mathbf{u}
        \right)
        \cdot
        \left(
            \mathbf{x^\prime} - \mathbf{x}
        \right)
    }
    {
        {\left|\mathbf{x^{\prime}}-\mathbf{x}\right|^2}
    }
    w
    \left(
            \frac
            {\left|\mathbf{x}-\mathbf{x^{\prime}}\right|}
            {r_e}
    \right)
    d\mathbf{x^{\prime}}.
\end{equation}
Equation \ref{eq:mpsdivvv} can be rewritten as:
\begin{equation}
\label{eq:mpsdiv}
\left\langle \nabla \cdot \mathbf{u}  \right\rangle \left(  \mathbf{x}\right)
\approx
-
\frac{d}{\left(r_e\right)^d A_0}
\int_{\mathbb{R}^d}
    \frac
    {
        \left(
            \mathbf{u^\prime} - \mathbf{u}
        \right)
        \cdot
        \left(
            \mathbf{x} - \mathbf{x^\prime}
        \right)
    }
    {
        {\left|\mathbf{x}-\mathbf{x^{\prime}}\right|}
    }
    \frac
    {
        1
    }
    {
        {\left|\mathbf{x}-\mathbf{x^{\prime}}\right|}
    }
    w
    \left(
            \frac
            {\left|\mathbf{x}-\mathbf{x^{\prime}}\right|}
            {r_e}
    \right)
    d\mathbf{x^{\prime}}.
\end{equation}
It is possible to compare it with the SPH expression of the divergence
(see section \ref{ss:appendix:grad} in the appendix).
As with the gradient, if expressions (\ref{eq:mpsdiv})
and (\ref{eq:contsphdiv}) were to coincide for every vector field,
necessarily the identity (\ref{eq:hreWw}) must hold.
The same argument used for the gradient allows us to conclude that
(\ref{eq:mpsdiv}) is a consistent approximation of the divergence. 
\subsection{Laplacian}
\label{ss:mpslapl}
The Laplacian in MPS for a scalar
function $\phi$ at a particle
with co-ordinates $\mathbf{x}_i$ is estimated (see e.g. \citep{youngyoon1999}) with the sum:
\begin{equation}
\label{eq:laplacian}
\langle \Delta \phi \rangle_i
=
\frac{2d}{\lambda n_0}
\sum_{j\neq i}
\left[
    \left(
        \phi_j - \phi_i
    \right)
    w
    \left(
            \frac
            {\left|\mathbf{x}_j-\mathbf{x}_i\right|}
            {r_e}
    \right)
\right],
\end{equation}
with
\begin{equation}
\label{eq:lambda1}
\lambda=
\frac
{
    \int_{\mathbb{R}^d}
    w
    \left(
            \frac
            {\left|\mathbf{x}\right|}
            {r_e}
    \right)
    \left|\mathbf{x}\right|^2
    \,d\mathbf{x}
}
{
    \int_{\mathbb{R}^d}
    w
    \left(
            \frac
            {\left|\mathbf{x}\right|}
            {r_e}
    \right)
    \,d\mathbf{x}
}.
\end{equation}
Let us define:
\begin{equation}
\label{eq:A_2}
A_2=
\int_{\mathbb{R}^d}
\left|\mathbf{q}\right|^2
w
\left(
        \left|\mathbf{q}\right|
\right)
\,d\mathbf{q}.
\end{equation}
As for $A_0$, $A_2$ only depends on the form of the kernel and does
not depend on the number of particles nor on
the specific value of $r_e$. With this notation, we have:
\begin{equation}
\label{eq:lambda2}
\lambda=
\frac{\left(r_e\right)^2\, A_2}{A_0}.
\end{equation}
Substituting $n_0$ in equation (\ref{eq:laplacian})
with its value from equation (\ref{eq:1overn0})
and interpreting the sum as the discrete approximation of an integral,
we get a continuum analogy to the MPS discrete Laplacian:
\begin{equation}
\label{eq:mpslapl}
\left\langle  \Delta \phi\right\rangle \left(  \mathbf{x}\right)
\approx
\frac{2d}{\left(r_e\right)^{d+2} A_2}
\int_{\mathbb{R}^d}
    \left[
        \phi(\mathbf{x}^{\prime}) - \phi(\mathbf{x})
    \right]
    w
    \left(
            \frac
            {\left|\mathbf{x}-\mathbf{x^{\prime}}\right|}
            {r_e}
    \right)
    d\mathbf{x^{\prime}}.
\end{equation}
Again, in order to check
whether this expression is a consistent approximation of
the Laplacian, we compare it with the SPH version
from \citet{Morris+etal:1997}(see section \ref{ss:app:lapl:morris}).
If expressions (\ref{eq:mpslapl}) and (\ref{eq:sphlaplmorris})
were to coincide for every function $\phi$,
necessarily, the following relation must hold:
\begin{equation}
\label{eq:hreWwlapl}
    -
    \frac
    {
        2
    }
    {
        h^{d+1}
    }
    \tilde{W}^{\prime}
    \left(
            \frac
            {\left|\mathbf{x}\right|}
            {h}
    \right)
=
\frac{2d}{\left(r_e\right)^{d+2} A_2}
    \left|\mathbf{x}\right|
    w
    \left(
            \frac
            {\left|\mathbf{x}\right|}
            {r_e}
    \right).
\end{equation}
Under the same assumptions used for the gradient, equation (\ref{eq:hreWwlapl})
can be written as:
\begin{equation}
\label{eq:Wwlapla}
    \tilde{W}^{\prime}(q)
=
-
\frac{d}{A_2}
    q\,w(q).
\end{equation}
Equation (\ref{eq:Wwlapla}) can be used to obtain the function $\tilde{W}$ involved
in the definition of the SPH kernel from the MPS weighting function $w$.
As before, in order for equation (\ref{eq:Wwlapla}) to give rise to a well defined
SPH kernel, we must check that (\ref{eq:kernelint1}) also holds in this case.
This turns out to be true since combining equations (\ref{eq:A_2}), (\ref{eq:Wwlapla}) and (\ref{eq:wgradw_d}),
we deduce that
\begin{displaymath}
\int_{\mathbb{R}^d}
    \tilde{W}(\left|\mathbf{q}\right|) d\mathbf{q}
=
-
\frac{1}{d}
\int_{\mathbb{R}^d}
\left|\mathbf{q}\right|
\tilde{W}^{\prime}(\left|\mathbf{q}\right|)d\mathbf{q}
=
\frac{1}{A_2}
\int_{\mathbb{R}^d}
\left|\mathbf{q}\right|^2
w(\left|\mathbf{q}\right|)d\mathbf{q}
=
1.
\end{displaymath}
\subsection{Summary of consistency order}
The consistency of the MPS gradient, divergence and Laplacian operators has been
established. The order of consistency of these approximations is 
obtained by pursuing the analogy with the well established SPH results described in
section \ref{ss:app:sphconsistency}:
\[
\left\langle \nabla_{\mathbf{x}}\phi\right\rangle =\nabla_{\mathbf{x}}\phi +\mathcal{O}\left(  \left(r_e\right)^{2}\right),
\qquad
\langle \nabla \cdot \mathbf{u} \rangle = \nabla \cdot \mathbf{u}+\mathcal{O}\left(  \left(r_e\right)^{2}\right),
\qquad
\left\langle \Delta \phi\right\rangle =\Delta\phi+\mathcal{O}\left(  \left(r_e\right)^{2}\right).
\]

%


\subsection{Relation between the MPS weighting functions and the SPH kernels}
\label{ss:mpssphkernelequiv}
It is remarkable that the relations between
the SPH kernel and MPS weighting functions
for the gradient and the Laplacian presented
in equations (\ref{eq:Ww}) and (\ref{eq:Wwlapla})
are different. The MPS method is somewhat equivalent
to SPH although different kernels for the gradient and the Laplacian are used.
This is not a common practice in SPH. However, some cases can be found in the literature,
as in \citep{laibe_price_2012a,laibe_price_2012b}, who
employ two types of kernels, one to interpolate densities
and buoyancies, and another one to calculate drag terms
in dusty gas simulations.

\begin{figure}[ht]
\centering
\includegraphics[width=0.9\textwidth]{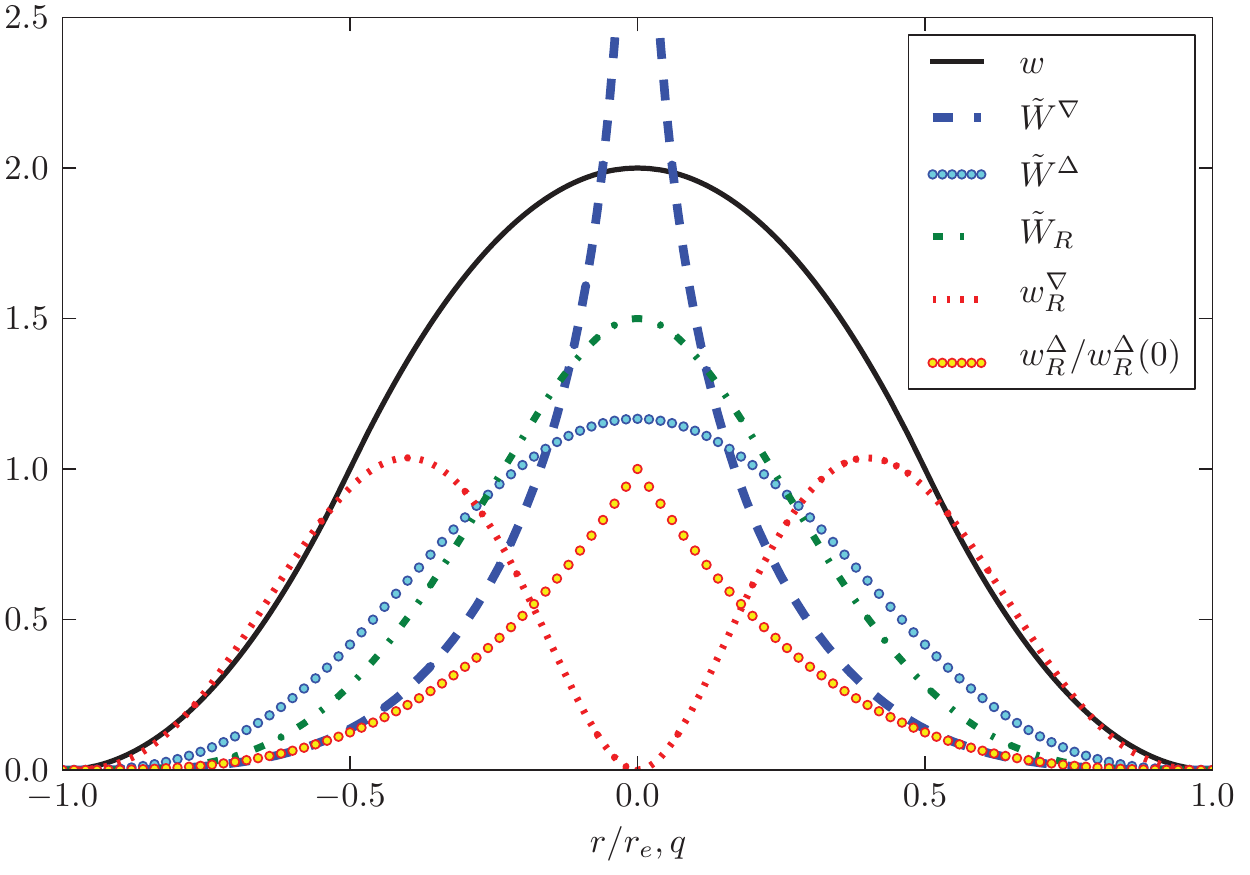}
\caption{MPS weighting function and SPH equivalent kernels}
\label{fig_dibujakernels}
\end{figure}

Let us build the equivalent SPH kernels of the MPS weighting function used in for instance
\citep{youngyoon1999}
\begin{equation}
w(r)=
\left\{
\begin{array}{cc}
  -(2r/r_e)^2+2, & 0\le r/r_e < 0.5, \\
  (2r/r_e - 2)^2, & 0.5< r/r_e \le 1, \\
  0, & 1 < r/r_e.  \\
\end{array}
\right.
\end{equation}
Let us remind the reader that $q=r/r_e$.
With this in mind, identity (\ref{eq:Ww}), referred to the
equivalence of the gradient operator between MPS and SPH,
is applied in order to obtain:
\begin{equation}
\label{Wwgradequiv}
\tilde{W}(q)=
-\frac{d}{A_0}
\int_0^q \frac{1}{s}w(s) ds + C.
\end{equation}
Let us use the notation $\tilde{W}_\nabla$ to denote the SPH kernel
obtained above. The constant $C$ is obtained by imposing $\tilde{W}_\nabla(1)=0$;
we therefore deduce:
\begin{equation}
\tilde{W}_\nabla(q)=-\frac{d}{A_0}
\left\{
\begin{array}{cc}
  -2q^2+2\log(q)+3+2\log(0.5),  & 0\le q \le 0.5, \\
  2q^2-8q+4\log(q)+6, & 0.5< q \le 1, \\
  0, & 1 < q.
\end{array}
\right.
\end{equation}
When for instance, $d=1$, one can easily check that $A_0=2$;
the graphs of $w$ and $\tilde{W_\nabla}$ are shown
in figure \ref{fig_dibujakernels}. Let us stress the fact
that $\tilde{W}_\nabla$ is a singular kernel, which is something seldom seen in the SPH literature.
The kernel is re-scaled by introducing the smoothing length as in equation (\ref{eq:sphkernel}):
\begin{equation}
\label{eq:sphmpsgrad}
W_\nabla\left(  \mathbf{x};h\right)  =\frac{1}{h^{d}}\tilde{W}_\nabla\left(  \left\vert
\frac{\mathbf{x}}{h}\right\vert \right).%
\end{equation}
Turning now to the Laplacian, we can apply identity (\ref{eq:Wwlapla}) to recover the
analytic expression of $\tilde{W}$ as
\begin{equation}
\tilde{W}(q)=
-\frac{d}{A_2}
\int_0^q {s}\,w(s) ds + C.
\end{equation}
Let us use the notation $\tilde{W_\Delta}$ for this kernel.
The constant $C$ is obtained by imposing $\tilde{W_\Delta}(1)=0$, thus obtaining:
\begin{equation}
\tilde{W_\Delta}(q)=-\frac{d}{A_2}
\left\{
\begin{array}{cc}
  -q^4+q^2 -7/24,  & 0\le q \le 0.5, \\
  q^4-8/3q^3+2q^2-1/3,\,\, & 0.5< q \le 1, \\
  0, & 1 < q .
\end{array}
\right.
\end{equation}
When $d=1$, one can easily check that $A_2=1/4$;
the graphs of $w$  and $\tilde{W_\Delta}$
are plotted in figure \ref{fig_dibujakernels}.  $\tilde{W_\Delta}$
is a non-singular kernel, a  compactly
supported class 2 piecewise polynomial function.
$W_\Delta$ is defined in a similar way as $W_\nabla$.
%
As already pointed out in section \ref{ss:mpslapl} it is relevant that the gradient and Laplacian SPH
derived kernels are different. It is interesting to note that the pressure effects,
which are controlled by the gradient value, are
modeled by using a singular kernel. The diffusive effects are on the other hand modeled by a
regular kernel, with zero slope at the origin, which is necessary, as demonstrated by \citet{violeau_2009}
to obtain correct dissipation values.

It is pertinent to compare both kernels with the fourth order Wendland one 
(order here refers to the differentiability order at the boundary of the compact support), 
which has become very popular among SPH practitioners due to its outstanding 
anti-clumping properties (see e.g. 
\citep{dehnen_aly_wendland_2012,Robinson_PhDthesis,macia_etal_spheric11_wendland,valizadeh_monaghan_pof2012}).
It can be appreciated in figure \ref{fig_dibujakernels} that the Wendland kernel $\tilde{W}_R$
is not as spiky as $\tilde{W}_\Delta$ and not as
flat as $\tilde{W}_\nabla$. Actually, it makes sense to obtain the equivalent MPS
weighting function for this Wendland kernel specially for the gradient since singular
kernels present problems when discretizing kernel integrals, as will be later discussed
in this paper. In 1D the expression for the Wendland kernel is:
\begin{equation}
\tilde{W_R}(q)=\frac{3}{32}
\left\{
\begin{array}{cc}
 (2-2\,q)^4\,(1+4\,q),  & 0\le q \le 1, \\
  0, & 1 < q.
 \end{array}
\right.
\end{equation}
Equation (\ref{eq:Ww}) can now be used to obtain an MPS weighting function from an SPH kernel:
\begin{equation}
\label{eq:wW_wendland}
w(q):=- \frac{q}{d} \tilde{W}^{\prime}(q).
\end{equation}
Notice that, from identity (\ref{eq:wgradw_d}), with this definition of $w$ one has $A_0=1$.
%

For the Wendland kernel in 1D,
%
\begin{equation}
\label{eq:wW_wendland2}
w_R^\nabla(q)=-q \tilde{W_R}^{\prime}(q)
= 30
\left\{
\begin{array}{cc}
 q^2-3\,q^3+3\,q^4-\,q^5,  & 0\le q \le 1, \\
  0, & 1 < q.
 \end{array}
\right.
\end{equation}
Its graph can be seen in figure \ref{fig_dibujakernels}. One salient aspect to highlight is that
it has a double hump and goes to zero at the origin, which makes it
radically different from any of the MPS weighting functions
documented in the literature (see e.g. \cite{koshizuka1996,youngyoon1999}). The reason for this
is out of the scope of the present paper but may be related to the different nature
of the pressure gradient model used in MPS (substraction of pressures in the interaction
of a pair of particles) and the one used in SPH (addition of pressures in the interaction
of a pair of particles).

The same can be done with the Laplacian equivalence (equation (\ref{eq:Wwlapla})) to obtain
the MPS weighting function:
\begin{equation}
\label{eq:Ww_wendland}
w(q):=- \frac{d}{q} \tilde{W}^{\prime}(q).
\end{equation}
Notice that with this definition of $w$ one has $A_2=1$ from identity (\ref{eq:wgradw_d}) in the appendix.
%

For the Wendland kernel in 1D:
%
\begin{equation}
\label{eq:wRDelta}
w_R^\Delta(q)=-\frac{1}{q} \tilde{W_R}^{\prime}(q)
= 30
\left\{
\begin{array}{cc}
 1-3\,q+3\,q^2-q^3,  & 0\le q \le 1, \\
  0, & 1 < q.
 \end{array}
\right.
\end{equation}
Its graph can be seen in figure \ref{fig_dibujakernels}. It has a non zero derivative at the origin
so it essentially lays between the original weighting function $w$ from \citep{youngyoon1999}
and the singular weighting function used by e.g. \citet{Tsukamoto_cheng_2011}.
%
%
%

\section{Boundary value problems}
\label{s:bc}
\subsection{General}
The fractional step method presented in section \ref{s:governing_equations} requires solving a
Poisson problem in each time-step. This Poisson problem is a partial differential
equation boundary value problem defined in a compact domain $\Omega$ which
involves the numerical approximation of the Laplacian operator for the left hand side.
Dirichlet or mixed Neumann and Dirichlet boundary conditions (BCs) can be considered in order
to obtain a unique solution.

In this section an MPS integral formulation
of this boundary problem is analyzed in order to unveil inconsistencies
present in the differential operators introduced in section \ref{s:diff_operators}
when applied to bounded domains. The problem is formulated in a general
way as
\begin{equation}
\label{eq:genlaplproblem}
\left\{
  \begin{array}{l}
  \Delta P(\mathbf{x}) = f(\mathbf{x}), \qquad \mathbf{x}\in \Omega,\\
  P(\mathbf{x}) = g(\mathbf{x})  \qquad \mathbf{x}\in \partial \Omega_1,\\
  \partial_\mathbf{n} P(\mathbf{x}) = h(\mathbf{x})  \qquad \mathbf{x}\in \partial \Omega_2,
    \end{array}
\right.
\end{equation}
assuming that $\partial \Omega = \partial \Omega_1 \cup \partial \Omega_2$ and that
$\mathbf{n}$ is the exterior unitary normal to $\partial \Omega$.

The MPS Laplacian operator in its integral formulation (equation (\ref{eq:mpslapl})) is now
modified considering that the integrals over $\mathbb{R}^d$ are restricted
to $\Omega$, which is where $P$ is defined:
\begin{equation}
\left\langle  \Delta P \right\rangle \left(  \mathbf{x}\right)
\approx
\frac{2d}{\left(r_e\right)^{d+2} A_2}
\int_{\Omega}
    \left[
        P(\mathbf{x}^{\prime}) - P(\mathbf{x})
    \right]
    w
    \left(
            \frac
            {\left|\mathbf{x}-\mathbf{x^{\prime}}\right|}
            {r_e}
    \right)
    d\mathbf{x^{\prime}}.
\end{equation}
%
\subsection{Zero Laplacian problem}
\label{ss:zerolapl_problem}
A zero Laplacian problem ($f(\mathbf{x})=0$) for the pressure is relevant in this context since it is
common that free-surface problems be driven by hydrostatic pressure
dominated fields, which are linear, meaning that they have a zero Laplacian.

Writing the Laplacian in its integral formulation (equation (\ref{eq:mpslapl})),
it must hold for every $\mathbf{x}\in \Omega$ that:
\begin{equation}
\label{eq:zerolapl}
\int_{\Omega}
    \left[
        P(\mathbf{x}^{\prime}) - P(\mathbf{x})
    \right]
    w
    \left(
            \frac
            {\left|\mathbf{x}-\mathbf{x^{\prime}}\right|}
            {r_e}
    \right)
    d\mathbf{x^{\prime}}=0.
\end{equation}
This is equivalent to:
\begin{equation}
\label{eq:zerolapllll}
P(\mathbf{x})
\int_{\Omega}
    w
    \left(
            \frac
            {\left|\mathbf{x}-\mathbf{x^{\prime}}\right|}
            {r_e}
    \right)
    d\mathbf{x^{\prime}}
=
\int_{\Omega}
    P(\mathbf{x}^{\prime})
    \,
    w
    \left(
            \frac
            {\left|\mathbf{x}-\mathbf{x^{\prime}}\right|}
            {r_e}
    \right)
    d\mathbf{x^{\prime}}.
\end{equation}
It turns out that the only solutions to equation (\ref{eq:zerolapllll})
are the constant ones, as it is next shown.

Start noticing that since $P$ is a continuous function
in $\overline{\Omega}$, it will reach
a maximum in this domain. Let us denote by $\mathbf{x}^*$
the point at which $P$ takes its maximum.
If $P$ is not identically constant, then this maximum must be strict.
Hence, there exist points $\mathbf{x^{\prime}}$ such that $\left|\mathbf{x}^*-\mathbf{x^{\prime}}\right|<r_e$
and $P(\mathbf{x^{\prime}})<P(\mathbf{x^*})$.
Since the weighting function $w$ is non-negative (see section \ref{s:weightf}), then one must have:
\begin{equation}
P(\mathbf{x}^*)
\int_{\Omega}
    w
    \left(
            \frac
            {\left|\mathbf{x}^*-\mathbf{x^{\prime}}\right|}
            {r_e}
    \right)
    d\mathbf{x^{\prime}}
>
\int_{\Omega}
    P(\mathbf{x}^{\prime})
    \,
    w
    \left(
            \frac
            {\left|\mathbf{x}^*-\mathbf{x^{\prime}}\right|}
            {r_e}
    \right)
    d\mathbf{x^{\prime}}.
\end{equation}
This clearly contradicts equation (\ref{eq:zerolapllll})
and therefore $P(\mathbf{x^{\prime}})=P(\mathbf{x^*}), \forall \mathbf{x^{\prime}}$.

This forces $P$ to be a constant function and thus
the integral approximation to the boundary value
problem (\ref{eq:genlaplproblem}) with $f(\mathbf{x})=0$
does not have a solution unless $g$ is constant.
In such case, that constant is the only solution to the problem.
A practical consequence of this to the analysis of the many-neighbor limit of the
discrete MPS formulation will be discussed in section \ref{sss:app:dirichlet}.
It must be noticed that this regime is not reached in practical applications since
the number of neighbors is limited by computational cost.
%
\subsection{Constant sign source Poisson problem}
We now consider a constant sign Laplacian problem for the pressure. This
problem is relevant in this context
since the divergence of the intermediate
velocity field in equation (\ref{eq:poissonMPS})
can maintain a constant sign for some intermediate flow fields.
This can happen e.g. due to an expansion or contraction
in free-surface flows, as in the standing wave case discussed
in section \ref{ss:standing_wave}.

Let us suppose that $f(\mathbf{x})$ in problem (\ref{eq:genlaplproblem}) has
a positive constant sign ($f(\mathbf{x})\ge 0$ for every $\mathbf{x}\in\Omega$
with $f(\mathbf{x^\prime})> 0$ for some  $\mathbf{x^\prime}\in\Omega$).
Writing the Laplacian in its integral formulation (equation (\ref{eq:mpslapl})),
one has that for every $\mathbf{x}\in \Omega$:
\begin{equation}
\label{eq:constsignlapl}
\int_{\Omega}
    \left[
        P(\mathbf{x}^{\prime}) - P(\mathbf{x})
    \right]
    w
    \left(
            \frac
            {\left|\mathbf{x}-\mathbf{x^{\prime}}\right|}
            {r_e}
    \right)
    d\mathbf{x^{\prime}}= \frac{\left(r_e\right)^{d+2} A_2}{2d} f(\mathbf{x}),
\end{equation}
which is equivalent to
\begin{equation}
\label{eq:poissonfxge0}
P(\mathbf{x})
\int_{\Omega}
    w
    \left(
            \frac
            {\left|\mathbf{x}-\mathbf{x^{\prime}}\right|}
            {r_e}
    \right)
    d\mathbf{x^{\prime}}
=
\int_{\Omega}
    P(\mathbf{x}^{\prime})
    \,
    w
    \left(
            \frac
            {\left|\mathbf{x}-\mathbf{x^{\prime}}\right|}
            {r_e}
    \right)
    d\mathbf{x^{\prime}}
    -
    \frac{\left(r_e\right)^{d+2} A_2}{2d} f(\mathbf{x}).
\end{equation}
Since $P$ is a continuous function in $\Omega$, it will reach
a maximum in this domain. Let us denote by $\mathbf{x}^*$
the point in which $P$ takes its maximum and assume it is
strict for $\mathbf{x}$  close to $\mathbf{x}^*$ . Since the
weighting function satisfies $w\ge 0$ (see section \ref{s:weightf}), then
\begin{align}
P(\mathbf{x}^*)
\int_{\Omega}
    w
    \left(
            \frac
            {\left|\mathbf{x}^*-\mathbf{x^{\prime}}\right|}
            {r_e}
    \right)
    d\mathbf{x^{\prime}}
& >
\int_{\Omega}
    P(\mathbf{x}^{\prime})
    \,
    w
    \left(
            \frac
            {\left|\mathbf{x}^*-\mathbf{x^{\prime}}\right|}
            {r_e}
    \right)
    d\mathbf{x^{\prime}} \nonumber \\
& \ge
\int_{\Omega}
    P(\mathbf{x}^{\prime})
    \,
    w
    \left(
            \frac
            {\left|\mathbf{x}^*-\mathbf{x^{\prime}}\right|}
            {r_e}
    \right)
    d\mathbf{x^{\prime}}
    -
    \frac{\left(r_e\right)^{d+2} A_2}{2d} f(\mathbf{x}^*).
\end{align}
Therefore, since $f$ is non-negative, there is no
$P$ verifying equation (\ref{eq:poissonfxge0}) at $\mathbf{x}^*$.
Hence, the integral approximation to the
boundary value problem (\ref{eq:genlaplproblem}) with $f(\mathbf{x})\ge 0$
does not have a solution. A similar reasoning can be done for a constant
negative sign using the minimum of $P$ instead of its maximum to
reach an identical conclusion. Similar considerations to those made
in previous section regarding the practical implications of this issue
apply here as well.
%
\subsection{Consistent definition of MPS operators}
\label{ss:consistent_def_mps_operators}
\subsubsection{General}
The performance of the MPS Laplacian operator
in its integral formulation has
been shown to present significant problems for simple boundary value problems
in compact domains. The impact of such problems at a discrete
level will later be discussed with practical applications and
will be addressed using the following corrected formulas for the MPS gradient, divergence
and Laplacian, which involve the computation of boundary integrals using the related
SPH kernels to each operator. The need for these boundary integrals arises
from the fact that the integral domain is not $\mathbb{R}^d$ but $\Omega$.

The formulas are implemented in the
MPS method by incorporating
boundary integral terms (see \citep{deleffe_etal_spheric09,Colagrossi2009,ferrand_etal_2012,Maciaetal_PTP_2012}
for an analogous SPH treatment) and by using the
SPH kernel - MPS weighting function
equivalences presented in section \ref{ss:mpssphkernelequiv}.

The formulas to be introduced in sections \ref{sss:correctmpsgrad}-\ref{sss:correctedmpslapl}
involve the following volume integrals of the equivalent SPH kernels (section \ref{ss:mpssphkernelequiv}):
\begin{align}
\label{eq:gammanablacont}
\Gamma^\nabla\left(\mathbf{x}\right)&=\int_{\Omega}  W_\nabla\left(\mathbf{x}-\mathbf{x}^{\prime} ; r_e\right)    d\mathbf{x}^{\prime},\\
\label{eq:gammaDeltacont}
\Gamma^\Delta\left(\mathbf{x}\right) &=\int_{\Omega} W_\Delta\left(\mathbf{x}-\mathbf{x}^{\prime} ; r_e\right)    d\mathbf{x}^{\prime}.
\end{align}
The reason for introducing these integrals will be explained in the following sections.
Let us mention for the moment that equation (\ref{eq:gammanablacont}) will play a role
in our approximation to first order differential operators whereas equation (\ref{eq:gammaDeltacont})
will be used in the approximation to the Laplacian.
\subsubsection{Corrected MPS gradient}
\label{sss:correctmpsgrad}
The proposed corrected formula for the MPS gradient is
\begin{align}
\label{eq:mpsgradcorrectedcontinuos}
\left\langle \nabla_\mathbf{x} \phi\right\rangle \left(  \mathbf{x}\right)
&
\approx
\frac{d}{\left(r_e\right)^d A_0\,\Gamma^\nabla\left(\mathbf{x}\right)}
\int_{\Omega}
    \frac
    {
        \phi(\mathbf{x^{\prime}}) - \phi(\mathbf{x})
    }
    {
        {\left|\mathbf{x^{\prime}}-\mathbf{x}\right|^2}
    }
    \left(
        \mathbf{x^\prime} - \mathbf{x}
    \right)
    \nonumber \\
&
    w
    \left(
            \frac
            {\left|\mathbf{x}-\mathbf{x^{\prime}}\right|}
            {r_e}
    \right)
    d\mathbf{x^{\prime}}
    +
    \frac{1}{\Gamma^\nabla\left(\mathbf{x}\right)}
    \int_{\partial \Omega}
         \left(
            \phi(\mathbf{x^{\prime}}) - \phi(\mathbf{x})
        \right)
        W_\nabla\left(\mathbf{x}-\mathbf{x}^{\prime} ; r_e\right)
        \,
        \mathbf{n}
        \,
    d\mathbf{x}^{\prime}.
\end{align}
The proposed correction is aimed at correcting the fact that weight function summations
may not be complete close to the boundaries. For this reason the term  $\Gamma^\nabla(\mathbf{x})$
is introduced in the denominator (it must be noticed
that  $\Gamma^\nabla(\mathbf{x})$ is equal to one when $\mathbf{x}$ is far from the boundary).
Once this correction has been performed, the boundary term appears naturally as a result
of applying the Green's identity. It allows to perform accurate integrations
close to the boundaries (it must be noticed that this boundary term
vanishes identically when it is applied to a particle that
is far from the boundary).

A nice feature of formula (\ref{eq:mpsgradcorrectedcontinuos}) is that it combines the use of the
MPS weighting function $w$ and the equivalent SPH kernel $W_\nabla$.
Its discrete formulation requires the discretization of the surface
integral across $\partial \Omega$. Therefore it is necessary to
explicitly define this surface. This is a tricky issue in fragmented
flows but since MPS requires solving a boundary problem for the
fractional step method, evaluating this boundary integral does not
introduce additional difficulties. A simple alternative is to discretize $\partial \Omega$
in a series of surface elements of area $S_j$ each one with a particle
at its centroid and with an exterior unitary normal $\mathbf{n}_j$.
The integral in equation (\ref{eq:mpsgradcorrectedcontinuos}) can therefore be
discretized as:
\begin{align}
\label{eq:mpsgradcorrecteddiscrete}
\left\langle \nabla \phi\right\rangle _i
&
=
\frac{d}{n_0\,\Gamma^\nabla_i}
\sum_{j\neq i}
\left[
    \frac
    {
        \phi_j - \phi_i
    }
    {
        {\left|\mathbf{x}_j-\mathbf{x}_i\right|^2}
    }
    \left(\mathbf{x}_j-\mathbf{x}_i\right)
    w
    \left(
            \frac
            {\left|\mathbf{x}_j-\mathbf{x}_i\right|}
            {r_e}
    \right)
\right]
    \nonumber \\
&
    +
    \frac{1}{\Gamma^\nabla_i}
    \sum_{\mathbf{x}_j\in \partial\Omega}
        \left(
            \phi_j - \phi_i
        \right)
        W_\nabla\left(\mathbf{x}_j-\mathbf{x}_i ; r_e\right)
        \cdot
        \mathbf{n}_j
        \,
        S_j.
\end{align}
The kernel integrals (\ref{eq:gammanablacont}) and (\ref{eq:gammaDeltacont})
also have to be discretized to enter into this expression.
A coherent MPS approach to these discretizations, taking into account
equation (\ref{eq:1overn0}) is
\begin{align}
\label{eq:gammanabla_discrete}
\Gamma^\nabla_i&=
            \frac{A_0\,\left(r_e\right)^d}{n_0}
            \,
            \sum_{j\neq i}
            W_\nabla\left(\mathbf{x}_j-\mathbf{x}_i ; h\right)\text{,}
            \\
\label{eq:gammaDelta_discrete}
\Gamma^\Delta_i&=
            \frac{A_0\,\left(r_e\right)^d}{n_0}
            \,
            \sum_{j}
            W_\Delta\left(\mathbf{x}_j-\mathbf{x}_i ; h\right).
\end{align}
\subsubsection{Corrected MPS divergence}
Formula (\ref{eq:mpsgradcorrectedcontinuos}) for the approximation of the partial derivatives
is easily applied to the definition of a corrected formula for the MPS divergence:
\begin{align}
\label{eq:mps_div_correctedcontinuos}
\left\langle \nabla \cdot  \mathbf{u} \right\rangle \left(  \mathbf{x}\right)
&
\approx
\frac{d}{\left(r_e\right)^d A_0\,\Gamma^\nabla\left(\mathbf{x}\right)}
\int_{\Omega}
    \frac
    {
        \left(\mathbf{u}^{\prime} - \mathbf{u}\right)
        \left(\mathbf{x^{\prime}} - \mathbf{x})\right)
    }
    {
        {\left|\mathbf{x^{\prime}}-\mathbf{x}\right|^2}
    }
    w
    \left(
            \frac
            {\left|\mathbf{x}-\mathbf{x^{\prime}}\right|}
            {r_e}
    \right)
    d\mathbf{x^{\prime}}
    \nonumber \\
&
    +
    \frac{1}{\Gamma^\nabla\left(\mathbf{x}\right)}
    \int_{\partial \Omega}
        \left(\mathbf{u}^{\prime} - \mathbf{u}\right)
        \cdot
        \mathbf{n}
        \,
        W_\nabla\left(\mathbf{x}-\mathbf{x}^{\prime} ; r_e\right)
    d\mathbf{x}^{\prime}.
\end{align}
This integral can be discretized analogously to the gradient one.
\subsubsection{Corrected MPS Laplacian}
\label{sss:correctedmpslapl}%
The proposed corrected formula for the MPS Laplacian is
\begin{align}
\label{eq:mpslapl_cont_corrected}
\left\langle  \Delta \phi\right\rangle \left(  \mathbf{x}\right)
&
\approx
\frac{2d}{\left(r_e\right)^{d+2} A_2 \,\Gamma^\Delta\left(\mathbf{x}\right)}
\int_{\Omega}
    \left[
        \phi(\mathbf{x}^{\prime}) - \phi(\mathbf{x})
    \right]
    w
    \left(
            \frac
            {\left|\mathbf{x}-\mathbf{x^{\prime}}\right|}
            {r_e}
    \right)
    d\mathbf{x^{\prime}}
    \nonumber \\
&
+
    \frac{2}{\Gamma^\Delta\left(\mathbf{x}\right)}
    \int_{\partial \Omega}
    \frac
    {
        \phi(\mathbf{x^{\prime}}) - \phi(\mathbf{x})
    }
    {
        {\left|\mathbf{x^{\prime}}-\mathbf{x}\right|^2}
    }
    \left(\mathbf{x^{\prime}}-\mathbf{x}\right)
    W_\Delta\left(\mathbf{x}-\mathbf{x}^{\prime} ; r_e\right)
    \,
    \mathbf{n}
    \,
    d\mathbf{x}^{\prime}.
\end{align}
Introducing the boundary terms and the normalization factor
$\Gamma^\Delta$ aims, analogously to the gradient and divergence
approximations, at correcting the incompleteness of weight function summations
and corresponding inaccuracies close to the boundaries \citep{Maciaetal_PTP_2012}.

The discrete version for this formula can be written as follows:
\begin{align}
\label{eq:mpslapl_discrete_corrected}
\langle \Delta \phi \rangle_i
&
=
\frac{2d}{\lambda n_0\,\Gamma^\Delta_i}
\sum_{j\neq i}
\left[
    \left(
        \phi_j - \phi_i
    \right)
    w
    \left(
            \frac
            {\left|\mathbf{x}_j-\mathbf{x}_i\right|}
            {r_e}
    \right)
\right]
\nonumber
\\
&
+
    \frac{2}{\Gamma^\Delta_i}
    \sum_{\mathbf{x}_j\in \partial\Omega}
    \frac
        {
            \phi_j - \phi_i
        }
        {
            {\left|\mathbf{x}_j-\mathbf{x}_i\right|^2}
        }
        \left(\mathbf{x}_j-\mathbf{x}_i\right)
        W_\Delta\left(\mathbf{x}_j-\mathbf{x}_i ; r_e\right)
        \cdot
        \mathbf{n}_j
        \,
        S_j.
\end{align}
In practice, the need for these corrections will be justified with some
applications presented in section \ref{s:app}. 
\section{Applications}
\label{s:app}
\subsection{General}
A series of analyses at the discrete level are now
presented with the aim to show some practical consequences
of the problems and proposed solutions documented in section \ref{s:bc}.

The corrected formulas presented in section
\ref{ss:consistent_def_mps_operators}
incorporate the kernel integrals (\ref{eq:gammanablacont}) and (\ref{eq:gammaDeltacont}).
These are evaluated using a sweep across the particles. The first
discrete application will be to analyze the rate of convergence
of those summations. We will focus on a one dimensional evenly-spaced
setup, fixing the kernel/weighting function support radius and varying
the resolution $\Delta x$ by increasing the number of particles.

Next, the accuracy and convergence
of the MPS gradient and Laplacian operators
will be compared with the proposed corrected versions.
The divergence is not discussed in 1D since it coincides with
the gradient.

A zero Laplacian boundary value problem in 1D with Dirichlet boundary
conditions is then proposed. Dirichlet BCs are very important in MPS since
in the Poisson equation for the pressure, the value of the pressure
at the free surface particles is given.

A constant sign source term 1D Poisson problem with mixed Dirichlet and
Neumann boundary conditions is then proposed.
This is a relevant problem from a practical application
point of view since solid boundary conditions
for the pressure are modeled
using a Neumann boundary condition.

Finally a two dimensional realistic flow consisting of
the attenuation of a viscous standing wave is proposed.
This flow has a Dirichlet
BC on the free surface and a Neumann BC on the bottom of
the tank; moreover, an analytic solution is available
in the literature.
\subsection{Kernel integrals}
\label{ss:kernel_integrals}
As discussed in section \ref{ss:consistent_def_mps_operators} and
apparent from the new proposed formulae definitions presented in that section,
the normalization factors $\Gamma^\nabla$ and $\Gamma^\Delta$
are a fundamental part of the new formulae. It is hence
important to discuss how accurate the evaluation of these factors is,
with the neighbors' summations (\ref{eq:gammanabla_discrete}) and (\ref{eq:gammaDelta_discrete}),
before presenting comparisons between the classical formulae and the new proposed ones.

In this and the next three sections one dimensional problems will be considered.
In all these examples the computational domain will be the closed interval
$[0,1]$. The interaction range $r_e$ is set as 0.1. The
limit $\delta x \rightarrow 0$, keeping $r_e$ fixed corresponds
to the integral approximation discussed in the previous sections.
This range of the parameters is not reached in practical applications due
to computational cost but is the natural one to be considered in order
to discuss consistency of the operators.

\begin{figure}[ht]
\centering
\includegraphics[width=0.495\textwidth]{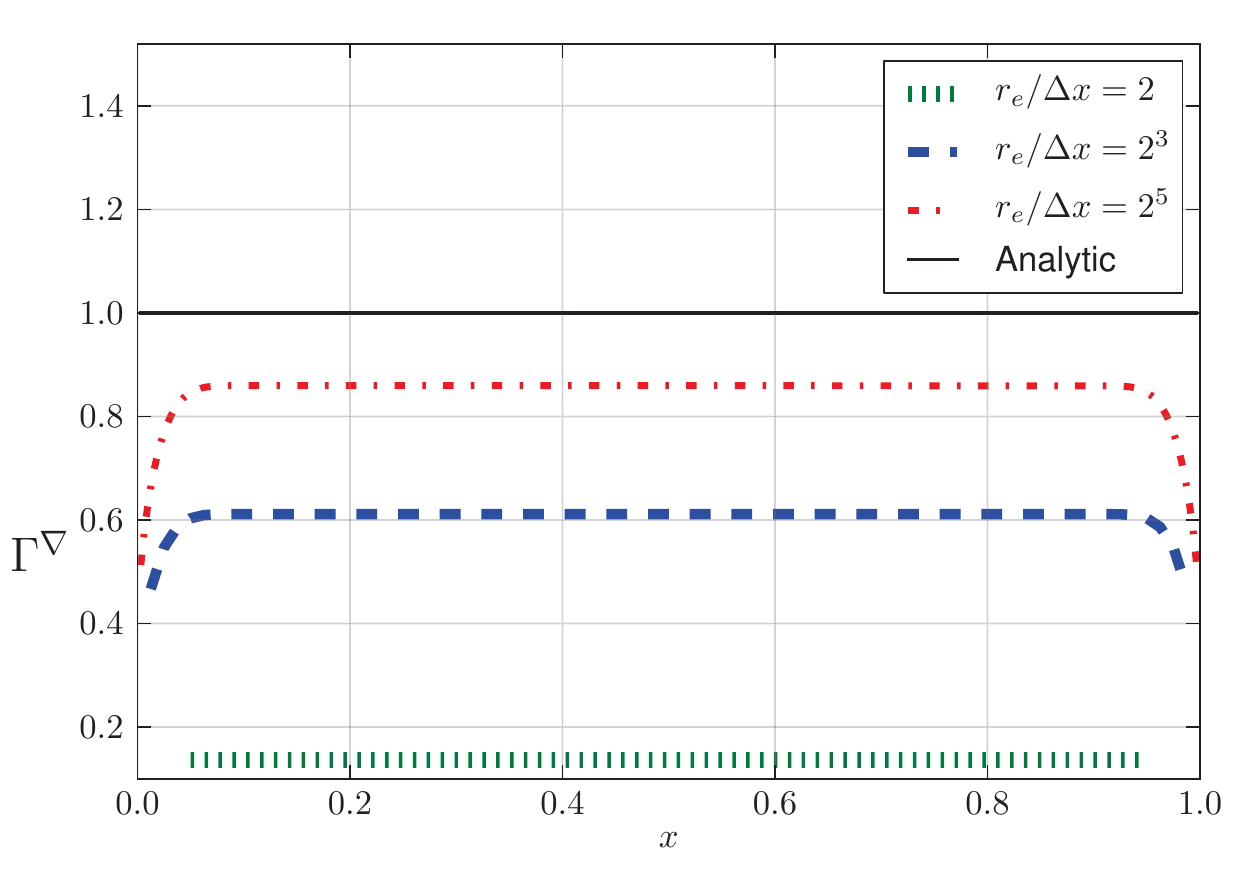}
\includegraphics[width=0.495\textwidth]{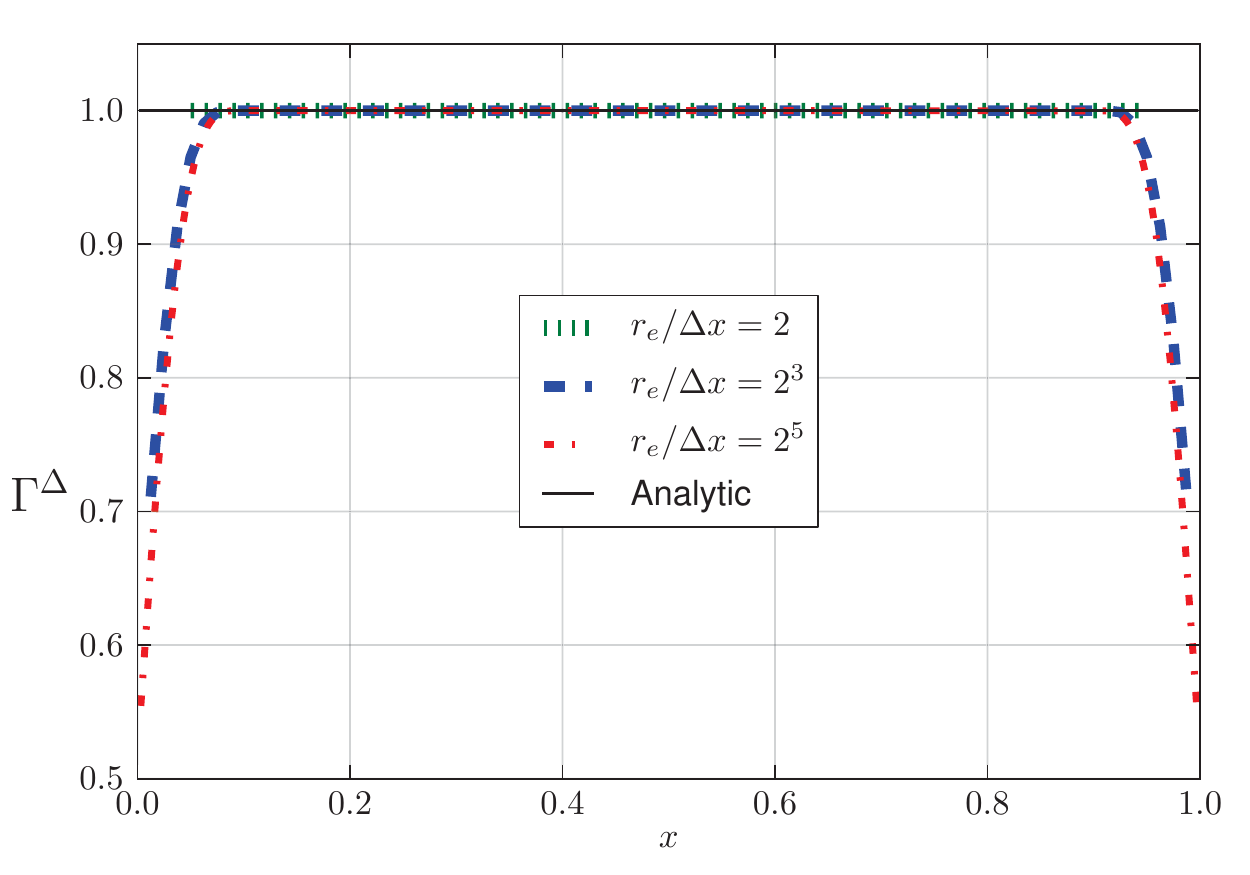}
\caption{Kernel summations computed with the MPS gradient derived SPH kernel $\Gamma^\nabla$ (left)
and with the MPS Laplacian derived SPH kernel $\Gamma^\Delta$ (right)}
\label{fig_evalua_gamma}
\end{figure}

In figure \ref{fig_evalua_gamma} (left) it can
be observed that for $\Gamma^\nabla$ the convergence to 1 is very
slow, the patent reason for this being the singular nature
of the kernel. The convergence to the exact values of the differential operators
which involve this renormalization
factor is therefore expected to be also slow. On the other hand,
the convergence for the $\Gamma^\Delta$ factor is good
(figure \ref{fig_evalua_gamma} right).
For $r_e / \Delta x = 2$ the exact value is obtained
which is a convenient property  to be later discussed.
%
\subsection{Gradient evaluation}
\label{ss:app:gradeva}
\begin{figure}[ht]
\centering
\includegraphics[width=0.495\textwidth]{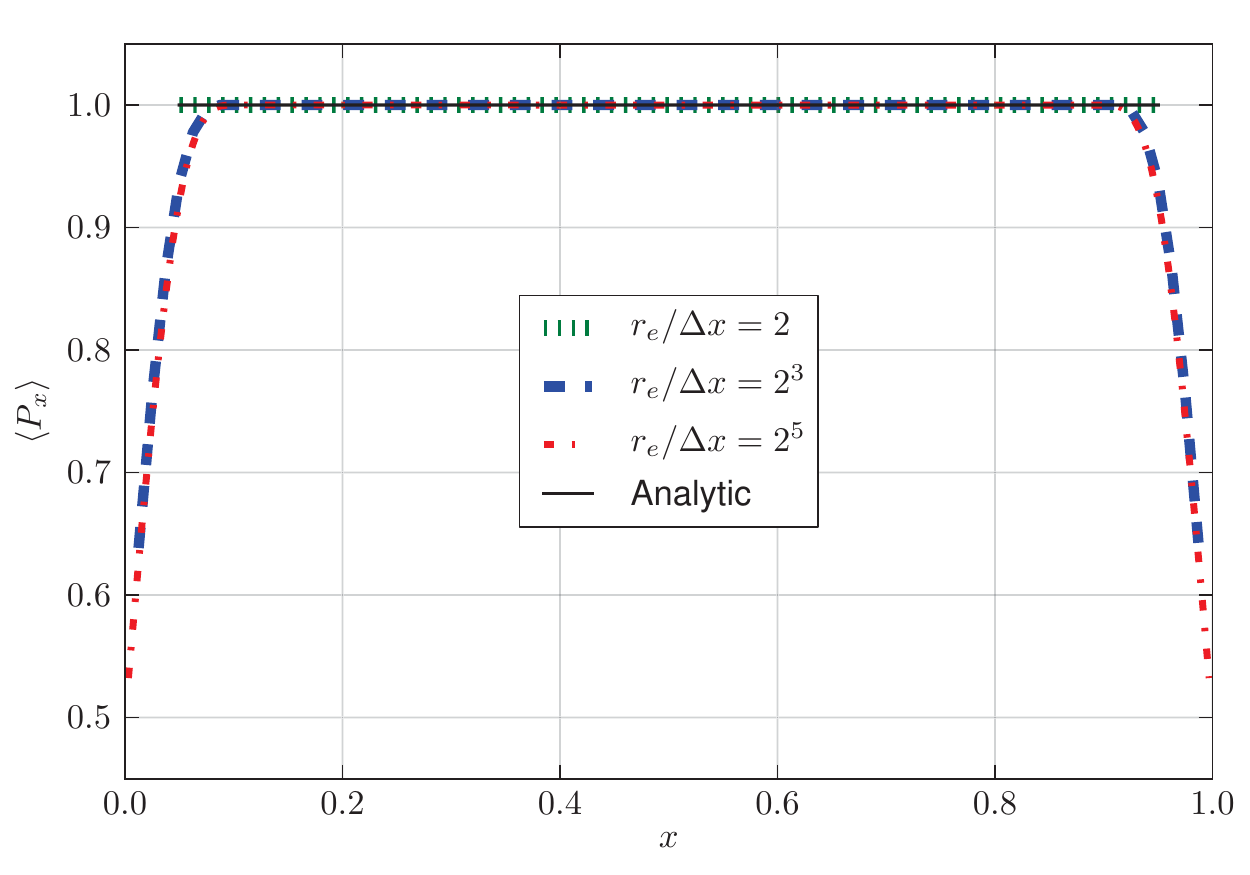}
\includegraphics[width=0.495\textwidth]{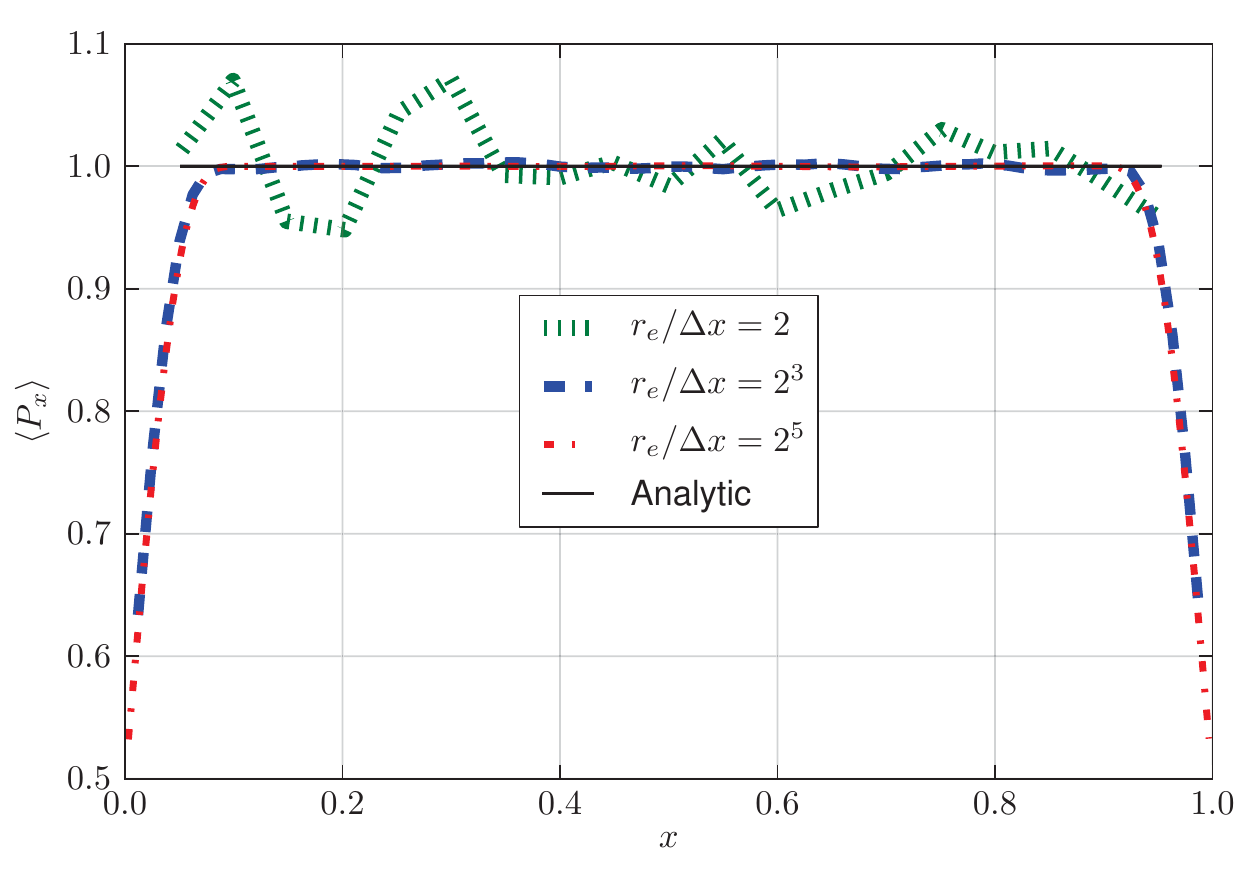}
\caption{Gradient evaluation (section \ref{ss:app:gradeva}) with MPS Gradient (equation \ref{eq:mpsgraddiscr}).
Regular lattice (left), disordered lattice (right).}
\label{fig_evalua_gradient_1d_q}
\end{figure}

Let us consider a linear field $P(x)= x$ for $x\in[0,1]$.
In figure \ref{fig_evalua_gradient_1d_q}
(left) it can be appreciated
that, keeping $r_e=0.1$ fixed,  the gradient approximation
provided by the classical MPS formula (\ref{eq:mpsgraddiscr})
converges to the exact value as
$\Delta x$ goes to zero ($r_e / \Delta x \rightarrow \infty$)
everywhere in the domain except at the boundaries,
where the values are halved.

It can be appreciated that for $r_e / \Delta x = 2$ the MPS evaluation of the gradient
returns the exact solution.
The solution for $r_e/\Delta x=2$ is exact because for evenly spaced particles, the
MPS operator is equivalent to a second order centered finite difference scheme.

This is very promising but in practical applications
the particles do not lie in a regular lattice.
Let us slightly modify the particles positions by displacing them
from the lattice with a random noise with maximum amplitude equal to $0.05\Delta x$.
With this new configuration the results presented in  figure
\ref{fig_evalua_gradient_1d_q} (right) are obtained.
These results indicate that the effect of this
low noise in lowering the accuracy of results is substantial.

It can be nonetheless appreciated that
increasing the number of neighbors becomes a means to increase accuracy even
for the disordered configuration.
On one hand, one could think that reducing $r_e$ for a fixed number of neighbors could solve these problems
but that may be in general not the case \citep{Quinlan_06,Amicarelli2011279}.

\begin{figure}[ht]
\centering
\includegraphics[width=0.495\textwidth]{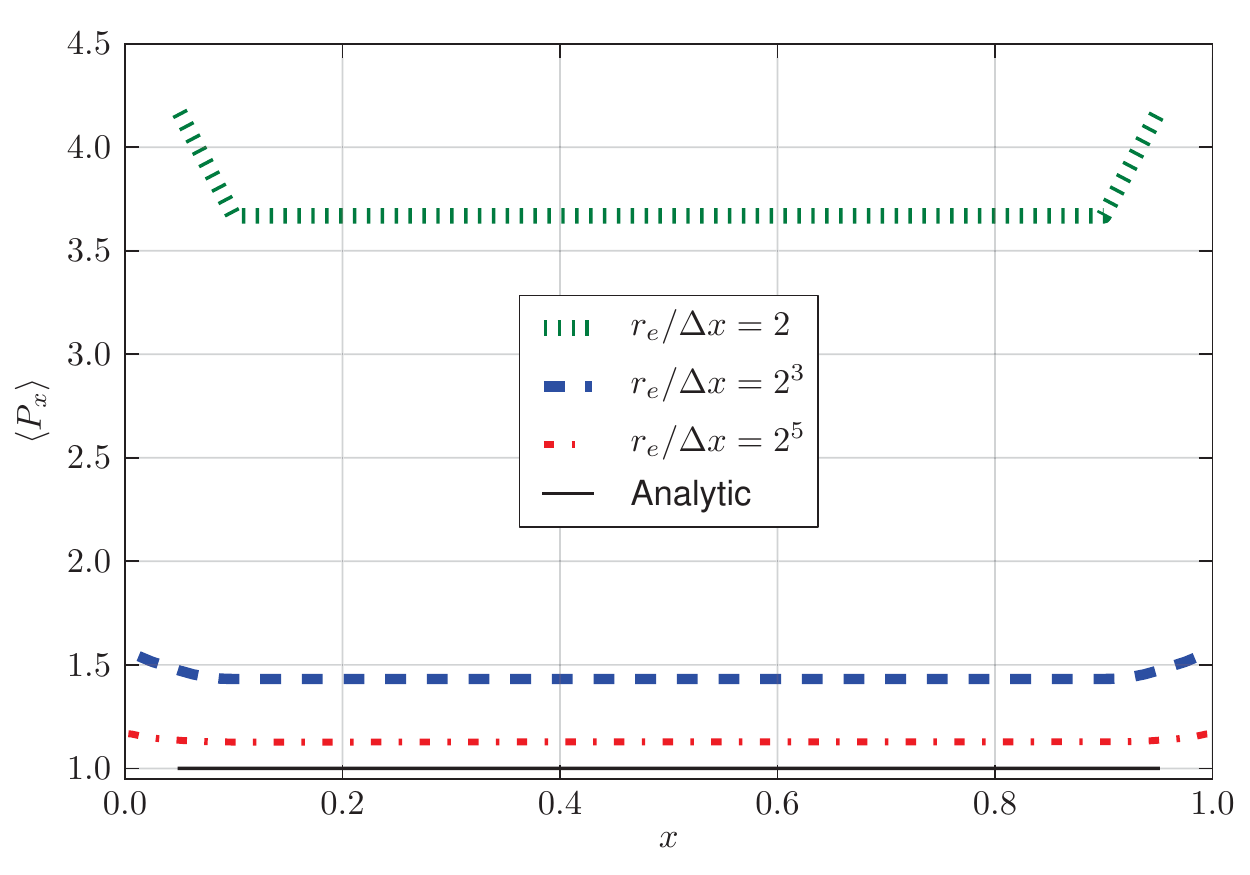}
\includegraphics[width=0.495\textwidth]{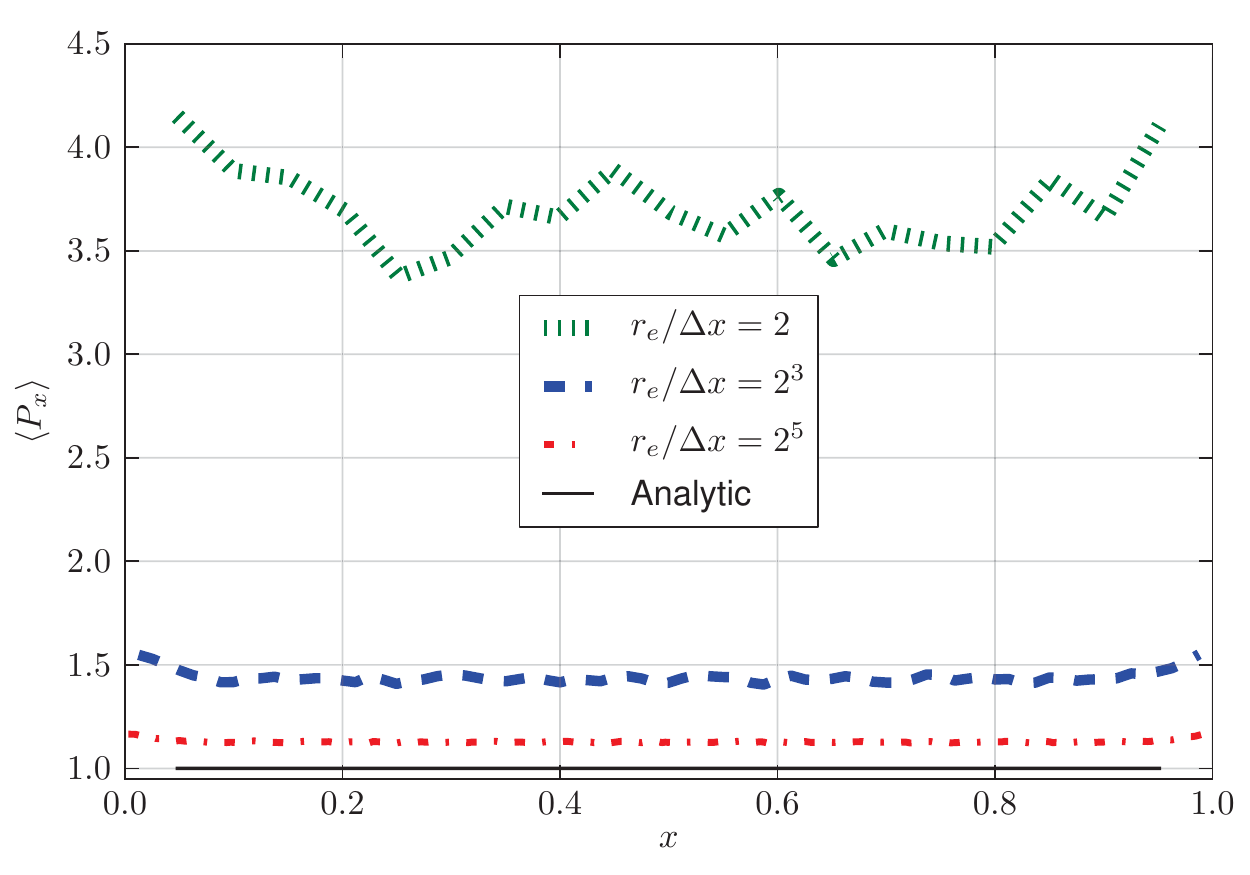}
\caption{Gradient evaluation (section \ref{ss:app:gradeva}) with corrected MPS Gradient (\ref{eq:mpsgradcorrecteddiscrete}).
Regular lattice (left), disordered lattice (right).}
\label{Xfig03left_evalua_gradient_1d_q_corrected}
\end{figure}

On the other hand, increasing the number of neighbors allows to have more accurate approximations in the
bulk of the domain but difficulties appear at the boundaries as can be seen in figures
 \ref{fig_evalua_gradient_1d_q} (left) and (right).
This lack of convergence to the exact solution close to the boundaries
can be overcome by correcting the gradient formula (\ref{eq:mpsgraddiscr}) as indicated in
section \ref{sss:correctmpsgrad}, using equation (\ref{eq:mpsgradcorrecteddiscrete}).
Its one dimensional form is:
\begin{align}
\label{eq:mpsgradcorrecteddiscrete1D}
\left\langle P_x \right\rangle _i
&
=
\frac{1}{n_0\,\Gamma^\nabla_i}
\sum_{j\neq i}
\left[
    \frac
    {
        P_j - P_i
    }
    {
        {\left|x_j-x_i\right|^2}
    }
    \left(x_j-x_i\right)
    w
    \left(
            \frac
            {\left|x_j-x_i\right|}
            {r_e}
    \right)
\right]
    \nonumber \\
&
    +
    \frac{1}{\Gamma^\nabla_i}
    \left[
         \left(
            P(1) - P_i
        \right)
        W_\nabla\left(1-x_i ; r_e\right)
        +
        \left(
            P_i-P(0)
        \right)
        W_\nabla\left( x_i ; r_e\right)
    \right].
\end{align}
Due to the singularity of the normalization factor for the gradient discussed
in section \ref{ss:kernel_integrals}, the convergence to the exact solutions for this formula is slow
(figure \ref{Xfig03left_evalua_gradient_1d_q_corrected} left) albeit clear, even close to the boundaries.
An MPS weighting function which would give rise
to an SPH regular kernel would be a good option to explore.
The convergence is not affected by disordering the particles as
can be appreciated in the left panel of figure \ref{Xfig03left_evalua_gradient_1d_q_corrected}.

Without entering into further consideration of
disordered particle configurations, it seems clear that the ratio $\Delta x / r_e$
must be small (thus implying a significant number of neighboring particles) in order to
obtain good interpolation properties for the first order differential operators (for both
general configurations and areas close to the boundaries).
In particular, this applies to the divergence operator, which is necessary
when evaluating the source term of the Poisson equation (\ref{eq:poissonMPS}) and
the pressure gradient for the momentum equation (\ref{Navier_Stokes_inc}).
%
%
%
\subsection{Laplacian evaluation}
\label{ss:app:lapleva}
\begin{figure}[ht]
\centering
\includegraphics[width=0.495\textwidth]{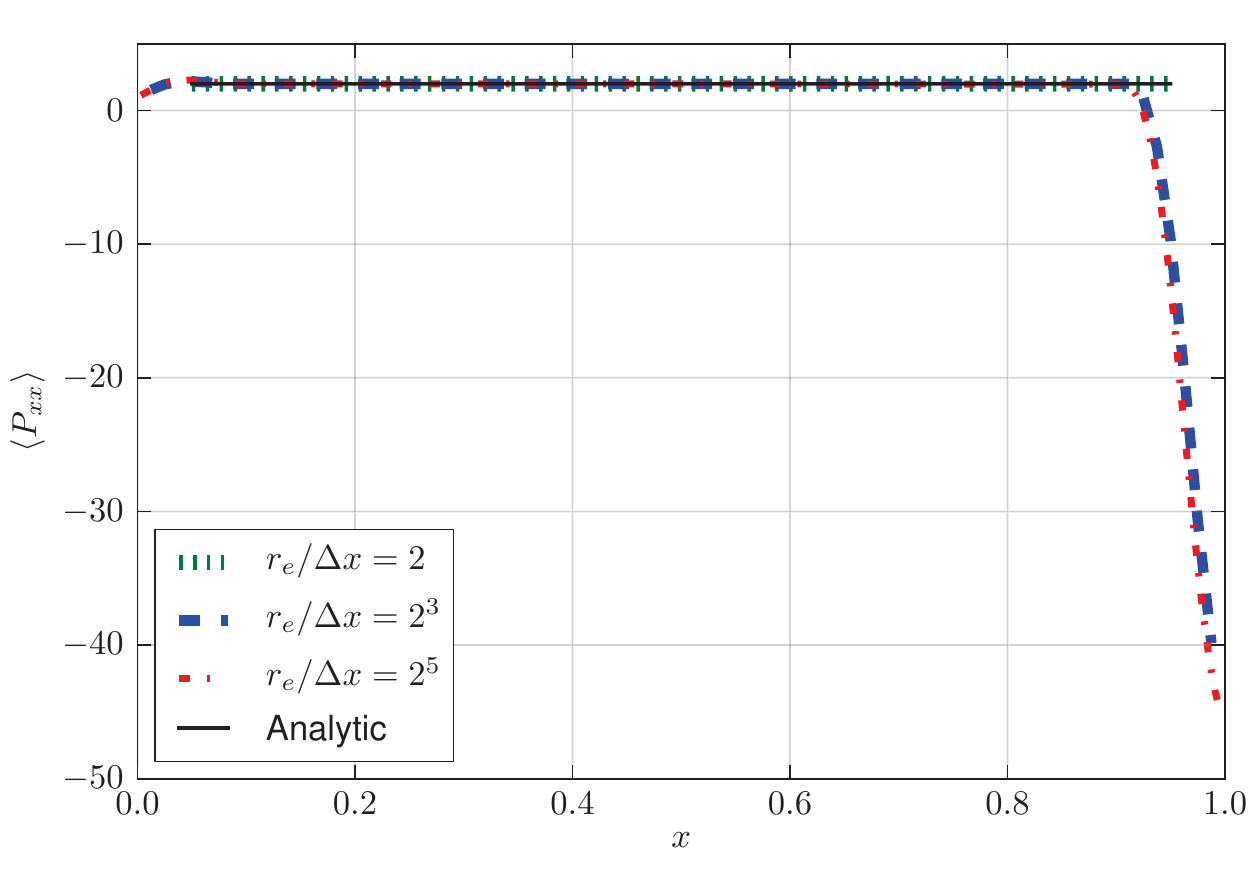}
\includegraphics[width=0.495\textwidth]{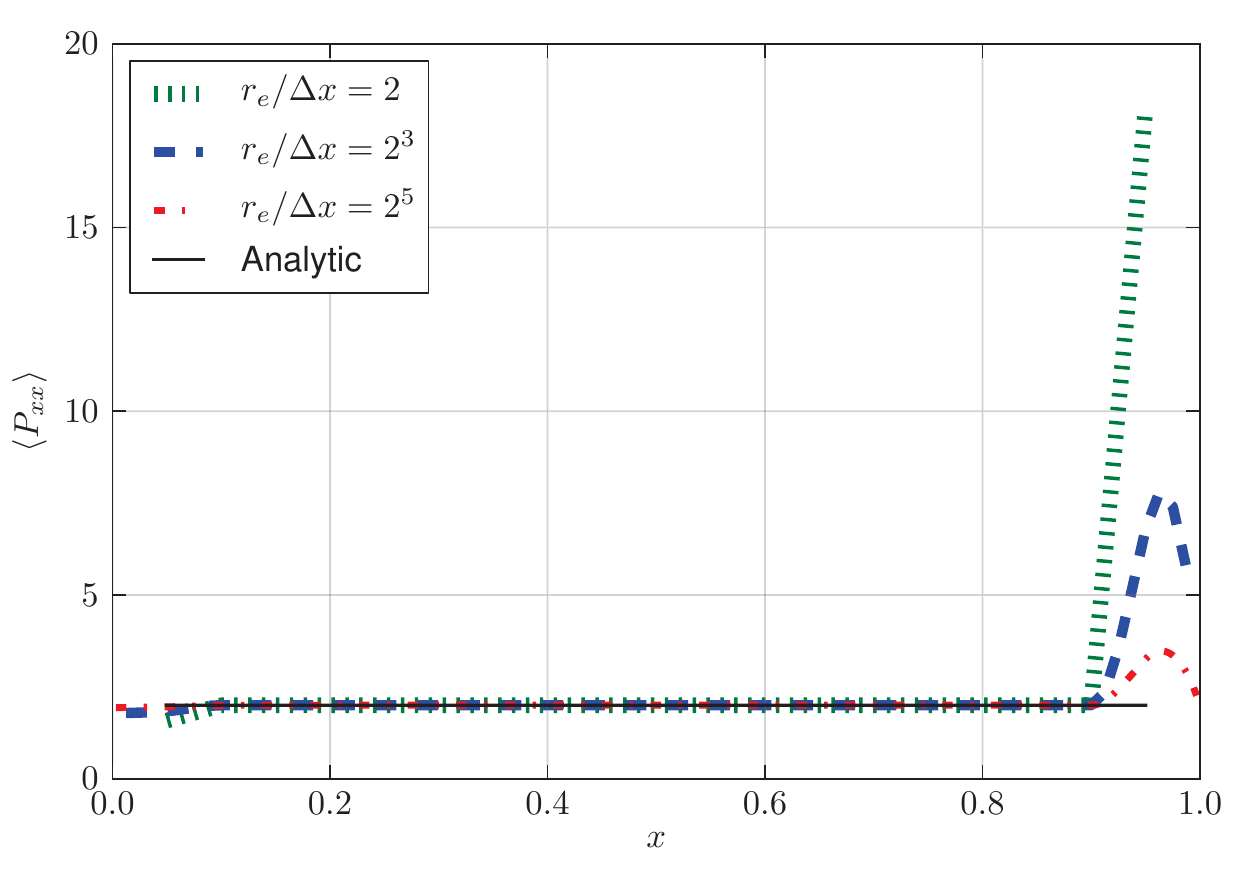}
\caption{Laplacian evaluation (section \ref{ss:app:lapleva}) with a regular lattice: MPS Laplacian (\ref{eq:laplacian}) (left), corrected MPS Laplacian (\ref{eq:mpslapl_discrete_corrected}) (right).}
\label{fig_evalua_laplacian_1d_q2}
\end{figure}

Let us now focus on a quadratic field $P(x)= x^2$ for $x\in[0,1]$.
In figure \ref{fig_evalua_laplacian_1d_q2}  (left) it can be appreciated
that, for  $r_e=0.1$, the approximation to the Laplacian
given by the classical MPS formulation (\ref{eq:laplacian})
converges to the exact value as
$\Delta x$ goes to zero everywhere in the interior of the domain.
This is no longer the case on the boundary points
where the operator becomes singular as $\Delta x / r_e$ goes to zero. Such singularity
has been subject of investigation in the SPH context in e.g. \citep{colagrossi_etal_pre2011,MaciaetalPTP}.
It can be overcome by correcting the formula as indicated in
section \ref{sss:correctedmpslapl} now using the equation which incorporates
the boundary terms with the SPH kernel. Its one dimensional form is:
\begin{align}
\label{eq:mps_lapl_correcteddiscrete1D}
\left\langle P_{xx} \right\rangle _i
&
=
\frac{2}{\lambda\,n_0\,\Gamma^\Delta_i}
\sum_{j\neq i}
\left[
    \left(
        P_j - P_i
    \right)
    w
    \left(
            \frac
            {\left|x_j-x_i\right|}
            {r_e}
    \right)
\right]
    \nonumber \\
&
    +
    \frac{2}{\Gamma^\Delta_i}
    \left[
        \frac
        {
            P(1) - P_i
        }
        {
             1 - x_i
        }
        W_\Delta\left(1-x_i ; r_e\right)
        -
        \frac
        {
            P_i - P(0)
        }
        {
             x_i
        }
        W_\Delta\left(x_i ; r_e\right)
    \right].
\end{align}
The convergence is clear (figure \ref{fig_evalua_laplacian_1d_q2} right)
but quite slow  due to the low order of the numerical derivatives
involved in the boundary terms contributions of
equation (\ref{eq:mpslapl_discrete_corrected}).

\begin{figure}[ht]
\centering
\includegraphics[width=0.9\textwidth]{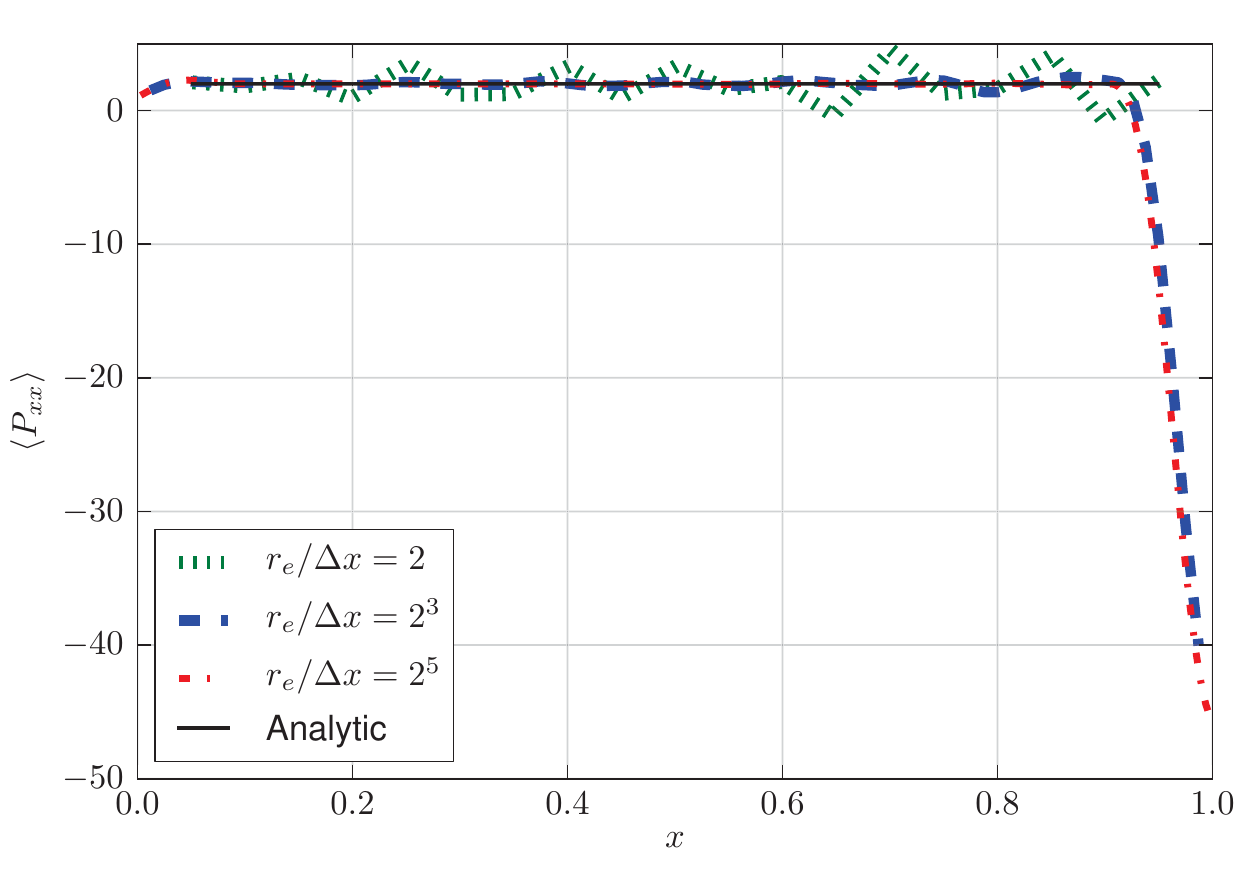}
\caption{Laplacian evaluation (section \ref{ss:app:lapleva}), MPS Laplacian (\ref{eq:laplacian}), noisy particle distribution.}
\label{fig_evalua_laplacian_1d_q2_noise}
\end{figure}

Coming back to figure \ref{fig_evalua_laplacian_1d_q2}, it can be appreciated that for
$r_e / \Delta x = 2$ the Laplacian is exact. This was also the case with the gradient
evaluation of section \ref{ss:app:gradeva}, but as in that case,
if the particles' positions are
displaced from the lattice with a
random noise of just $0.05\Delta x$ maximum amplitude, figure
\ref{fig_evalua_laplacian_1d_q2_noise} is obtained. This figure shows
that the effect of this small disorder in the accuracy of results for small
number of neighbors is substantial, and a similar discussion as that of
section \ref{ss:app:gradeva} applies here.

\subsection{Boundary value problems}
\subsubsection{1D zero Laplacian: Dirichlet boundary conditions}
\label{sss:app:dirichlet}
A 1D zero Laplacian  problem whose exact solution is $P(x)=x$ is proposed:
\begin{equation}
\label{1d_dirichlet}
\left\{
  \begin{array}{l}
  P^{\prime{}\prime{}}(x) = 0, \qquad x\in[0,1],\\
  P(0) = 0, \\
  P(1) = 1.
    \end{array}
\right.
\end{equation}
The domain $[0,1]$ is discretized by fixing the cut-off radius $r_e$ and
choosing the particle distance $\Delta x$ as a fraction of $r_e$. The MPS
weighting function from \citep{youngyoon1999} is used.

Equation (\ref{eq:laplacian}) is used in the MPS method to compute the discrete
Laplacian. Dirichlet boundary conditions are implemented in MPS (see i.e.
\citep{koshizuka1998,Tsukamoto_cheng_2011}) by identifying the particles defining the boundaries and then
applying to those particles the corresponding value of the pressure.
Their contribution to the Laplacian pass to the right hand side of the
linear system that results from discretizing the boundary value problem.

\begin{figure}[ht]
\centering
\includegraphics[width=0.9\textwidth]{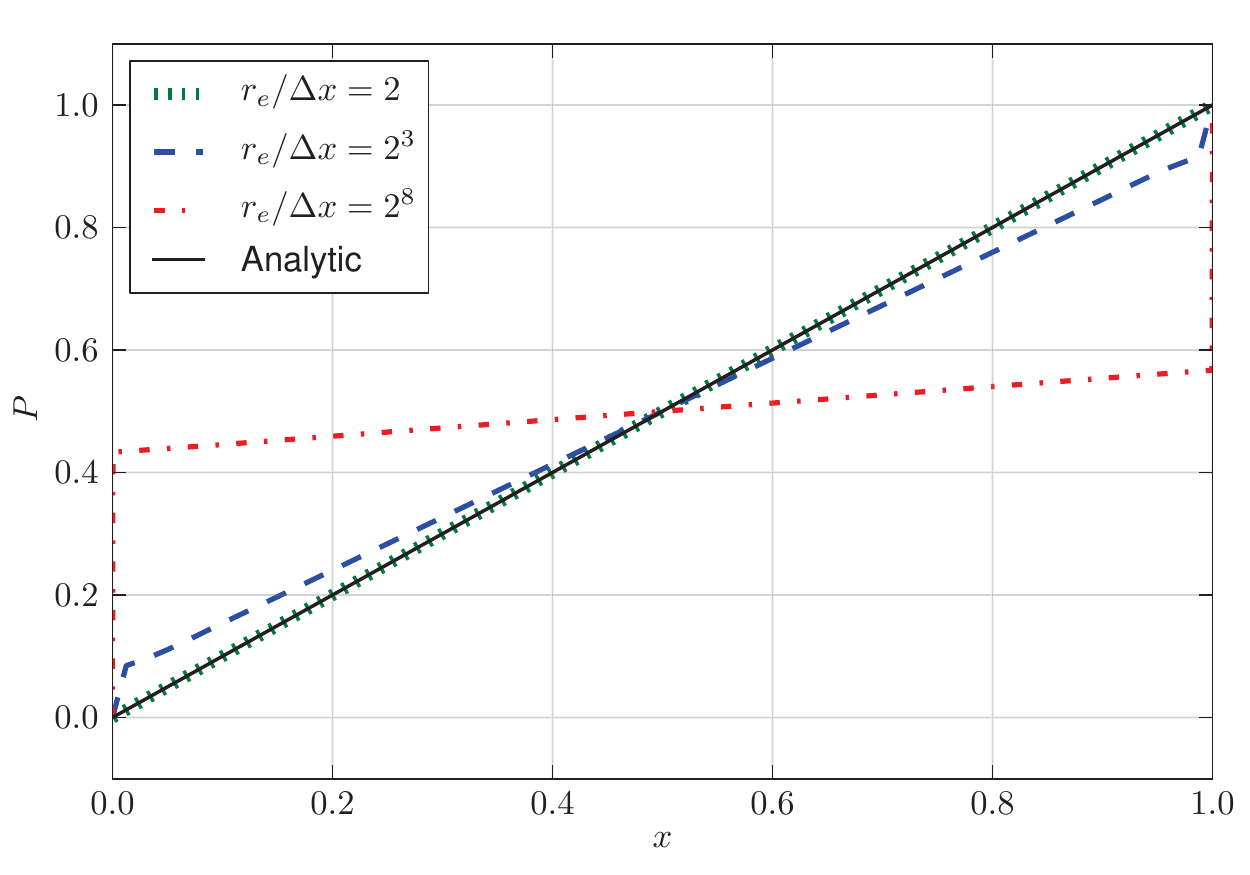}
\caption{Solution of the boundary value problem (\ref{1d_dirichlet}), MPS Laplacian (\ref{eq:laplacian}).}
\label{fig_poisson_1d_dirichlet}
\end{figure}

Assuming $r_e=0.1$ and evenly spaced particles,
the results for different values of the ratio $\Delta x/r_e$
are presented in figure \ref{fig_poisson_1d_dirichlet}.
It is clear that, as we tend to the continuum ($\Delta x \rightarrow 0$),
the solution does not converge to the analytical one $(p=x)$.
It actually becomes discontinuous and tends, as a consequence of the lack of consistency of the integral operator
discussed in section \ref{ss:zerolapl_problem}, to a constant function
in the interior of the domain.

The solution for $r_e/\Delta x=2$ is exact because for evenly spaced particles, the
MPS operator is equivalent to a second order centered finite difference scheme.
Nonetheless, the need for a larger number of neighboring particles lies in
the low interpolation properties of the gradient operator and Laplacian operator with
disordered particles discussed in sections \ref{ss:app:gradeva}
 and \ref{ss:app:lapleva} respectively.

\begin{figure}[ht]
\centering
\includegraphics[width=0.495\textwidth]{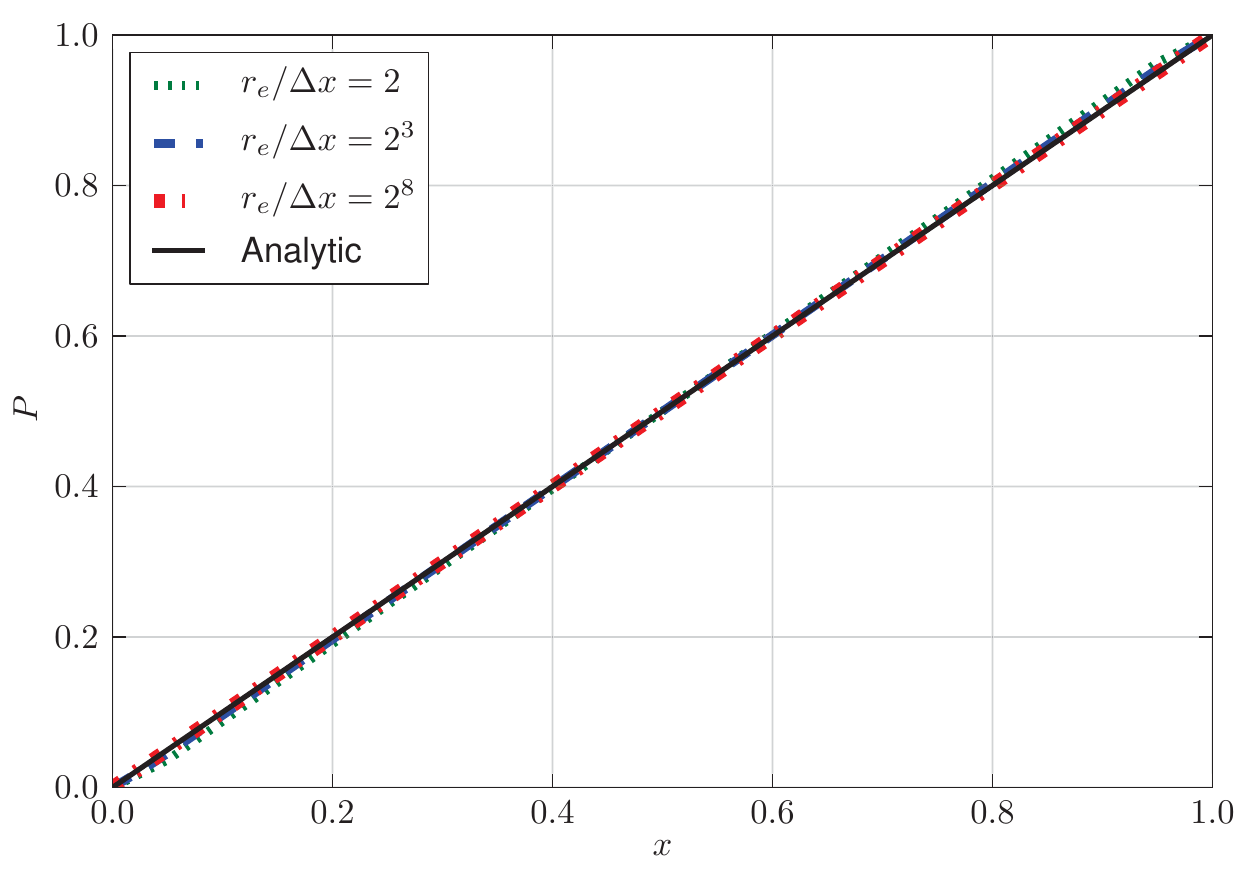}
\includegraphics[width=0.495\textwidth]{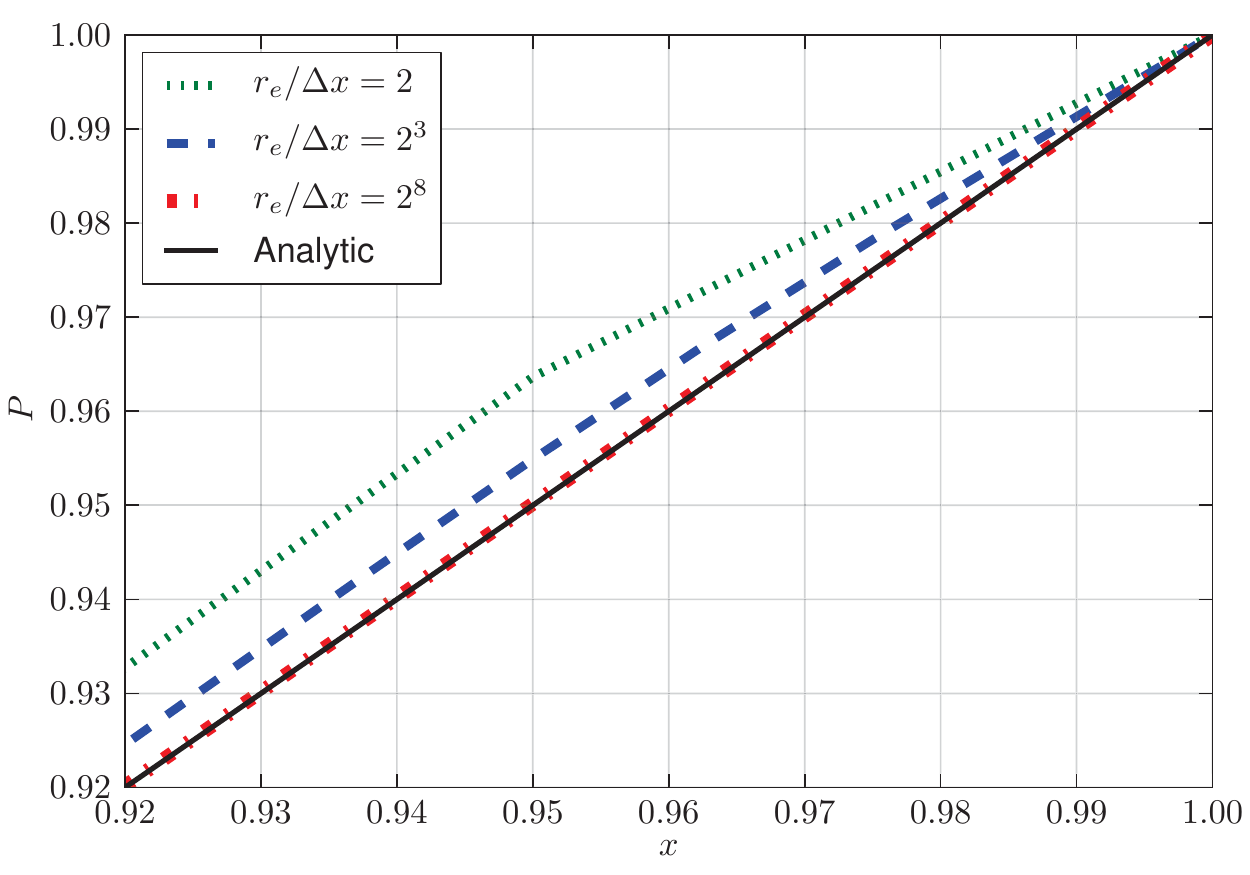}
\caption{Solution of the boundary value problem (\ref{1d_dirichlet}), corrected MPS Laplacian (\ref{eq:mpslapl_discrete_corrected}) (right, zoom).}
\label{fig_poisson1d_dirichlet_corrected}
\end{figure}

If the corrected Laplacian operator (\ref{eq:mpslapl_discrete_corrected}) is applied,
the convergence to the exact solution is clear across the domain
(figure \ref{fig_poisson1d_dirichlet_corrected} left and zoom in right).
 %
%
\subsubsection{1D constant sign source: Dirichlet and Neumann boundary conditions}
\label{s:app:dirichlet_neumann}
A 1D Poisson problem whose exact solution is $P(x)=x^2$ and which uses mixed
Neumann and Dirichlet boundary conditions is proposed:
\begin{equation}
\label{1d_neumann}
\left\{
  \begin{array}{l}
  P^{\prime{}\prime{}}(x) = 2, \qquad x\in[0,1],\\
  P^{\prime}(0) = 0, \\
  P(1) = 1.
    \end{array}
\right.
\end{equation}
The domain $[0,1]$ is discretized by fixing the cut-off radius $r_e$ and
choosing the particle distance $\Delta x$ as a fraction of $r_e$. The MPS
weighting function from \citep{youngyoon1999} is used.

\begin{figure}[ht]
\centering
\includegraphics[width=0.85\textwidth]{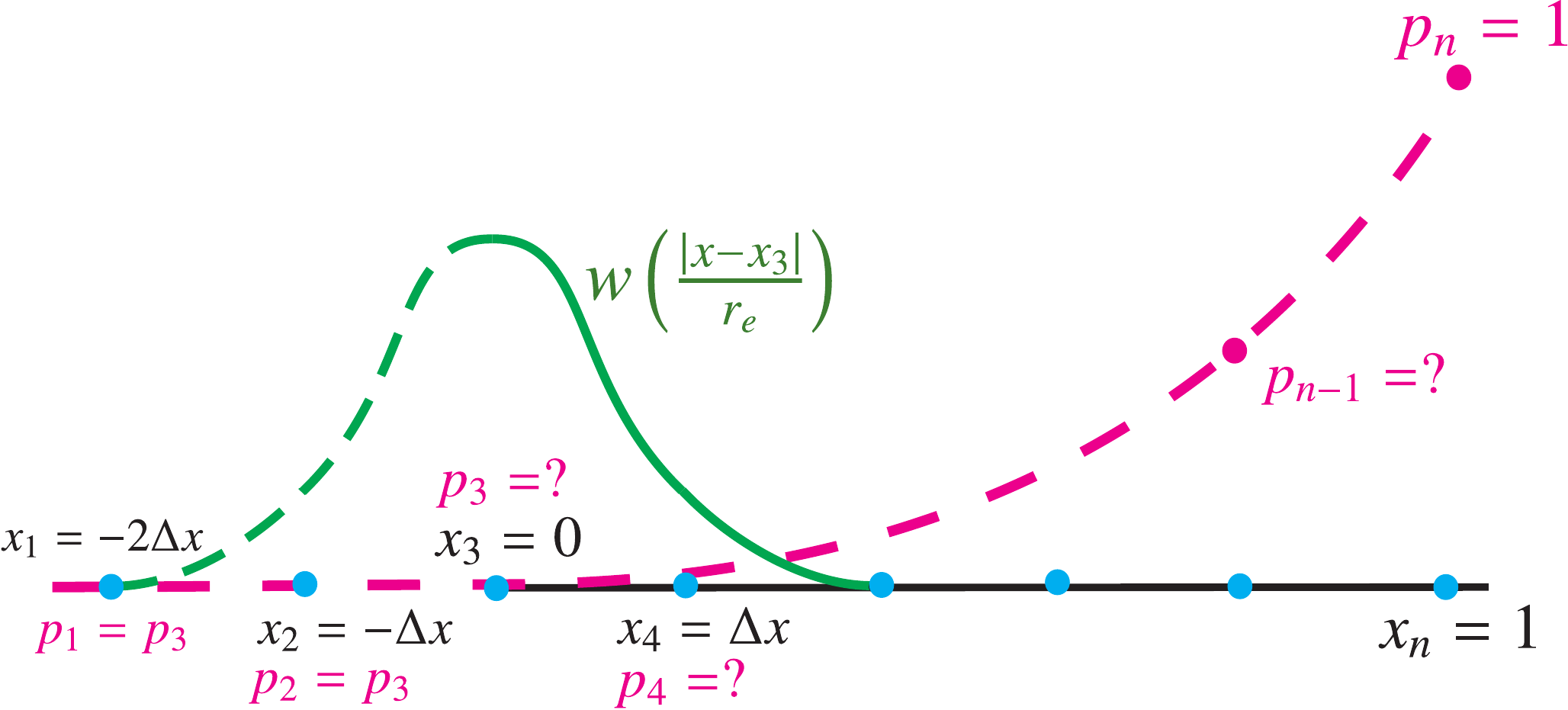}
\caption{Boundary value problem (\ref{1d_neumann}) discretization example sketch, with $r_e/\Delta x=2$.}
\label{fig_neumann1d_discretedomain}
\end{figure}

Equation (\ref{eq:laplacian}) is used in the MPS scheme to compute the discrete
Laplacian. The Dirichlet boundary conditions implementation has been discussed
in section \ref{sss:app:dirichlet}.
The Neumann boundary conditions are implemented by creating a set of dummy
particles in the region with $x<0$.
The pressure of all these dummy particles is taken equal
to the pressure of the particle
on the left boundary, $(x=0)$, which is in itself an unknown.
The number of dummy particles is chosen in such a way that the number of neighbors remains constant \citep{Tsukamoto_cheng_2011}.
In figure \ref{fig_neumann1d_discretedomain} a sketch of the setup is presented.
%

\begin{figure}[ht]
\centering
\includegraphics[width=0.95\textwidth]{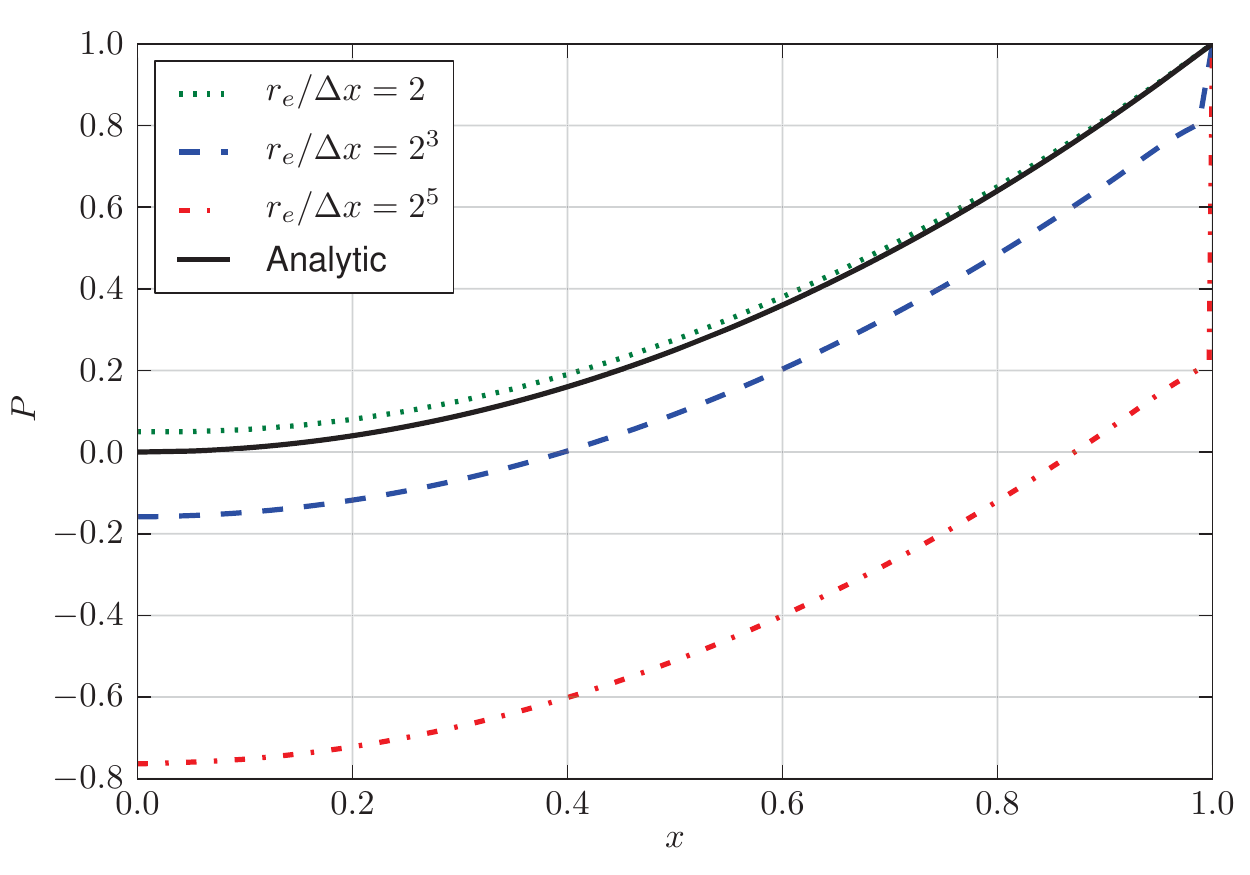}
\caption{Solution of the boundary value problem (\ref{1d_neumann}), MPS Laplacian (\ref{eq:laplacian}).}
\label{fig_poisson_1d_dirichlet_neumann}
\end{figure}

Assuming $r_e=0.1$ the results for different values of the ratio $r_e/\Delta x$
are presented in figure \ref{fig_poisson_1d_dirichlet_neumann}.
It is clear that as we tend to the continuum ($\Delta x \rightarrow 0$),
the solution does not converge to the analytical one $(p(x)=x^2)$. It does not either converge to a
continuous solution of the integral formulation (\ref{eq:constsignlapl}),
becoming discontinuous at $(x=1)$.

As in the previous section, the solution for $r_e/\Delta x=2$ is exact for similar reasons.

\begin{figure}[ht]
\centering
\includegraphics[width=0.495\textwidth]{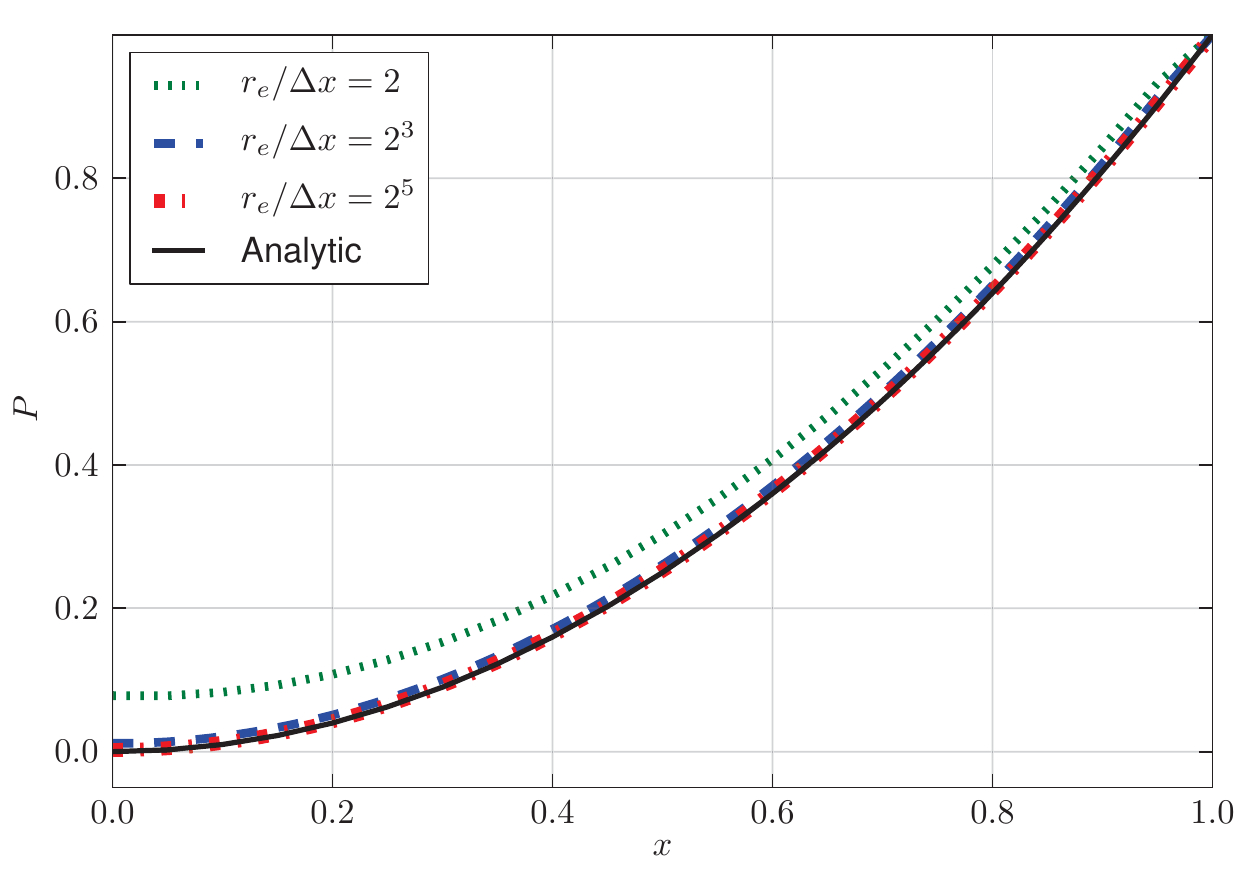}
\includegraphics[width=0.495\textwidth]{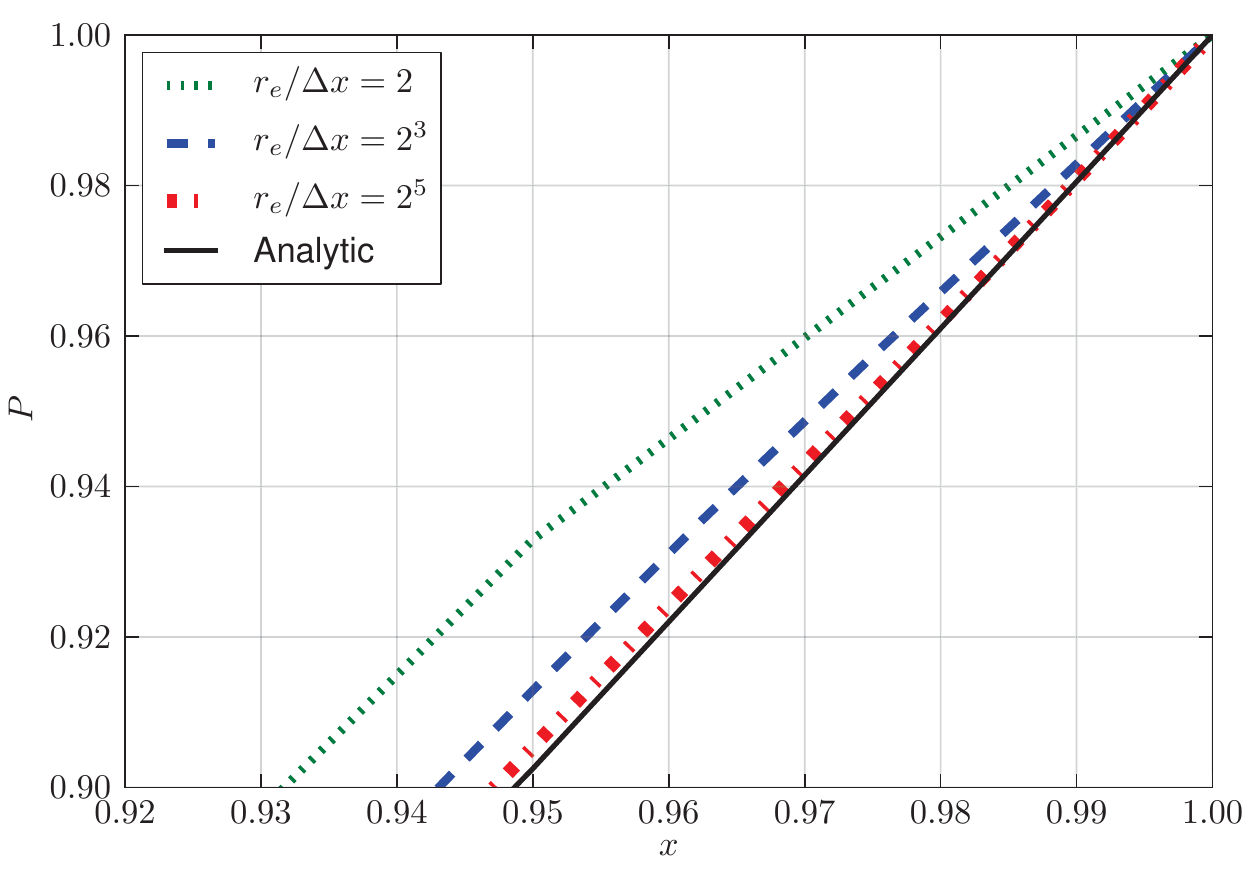}
\caption{Solution of the boundary value problem (\ref{1d_neumann}), corrected MPS Laplacian (\ref{eq:mpslapl_discrete_corrected}) (right, zoom).}
\label{fig_poisson1d_neumann_corrected}
\end{figure}

If the corrected Laplacian operator (\ref{eq:mpslapl_discrete_corrected}) is applied,
the exact solution is recovered for all the
discretizations (figure \ref{fig_poisson1d_neumann_corrected} left and zoom in right).

%
%
%
\subsubsection{Symmetrized corrected Laplacian}
\label{s:app:symmetrized}
\begin{figure}[ht]
\centering
\includegraphics[width=0.495\textwidth]{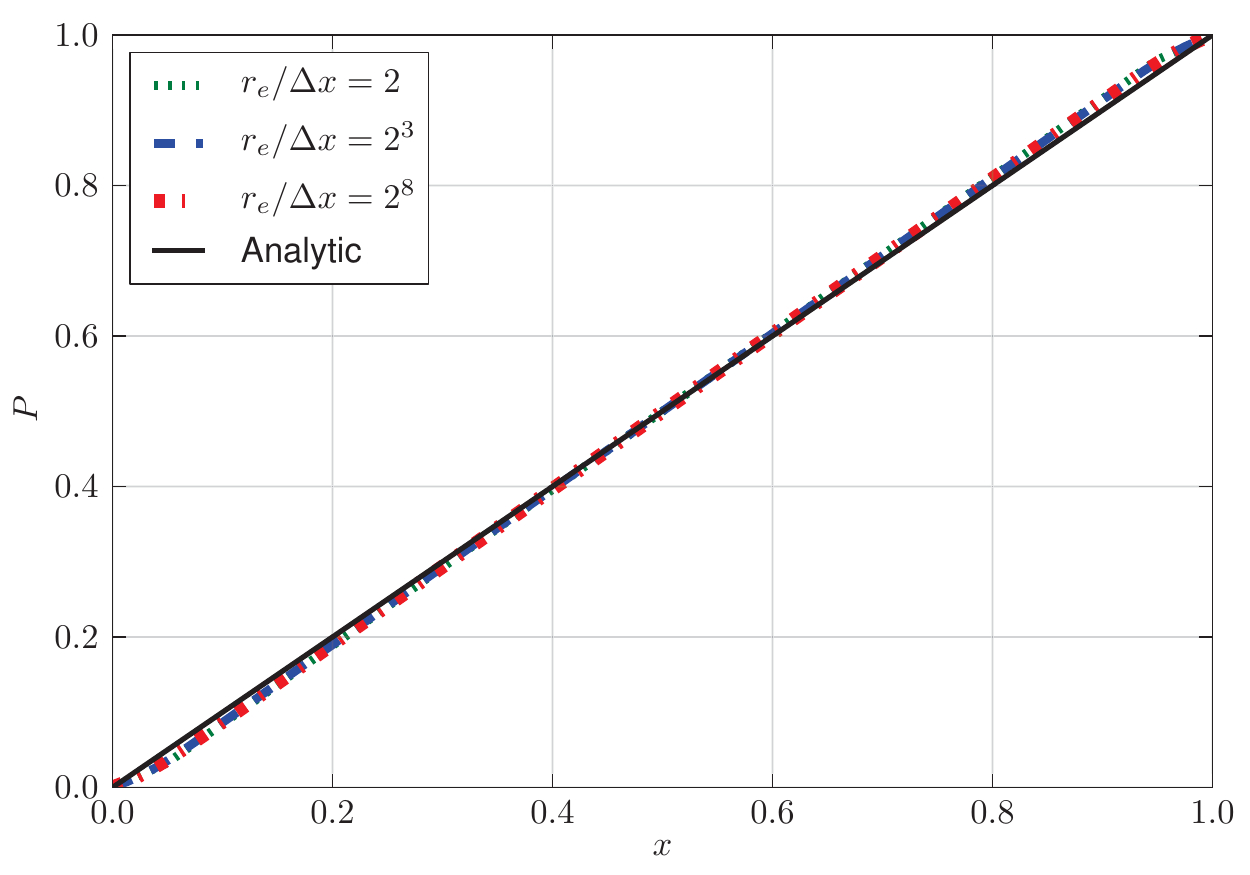}
\includegraphics[width=0.495\textwidth]{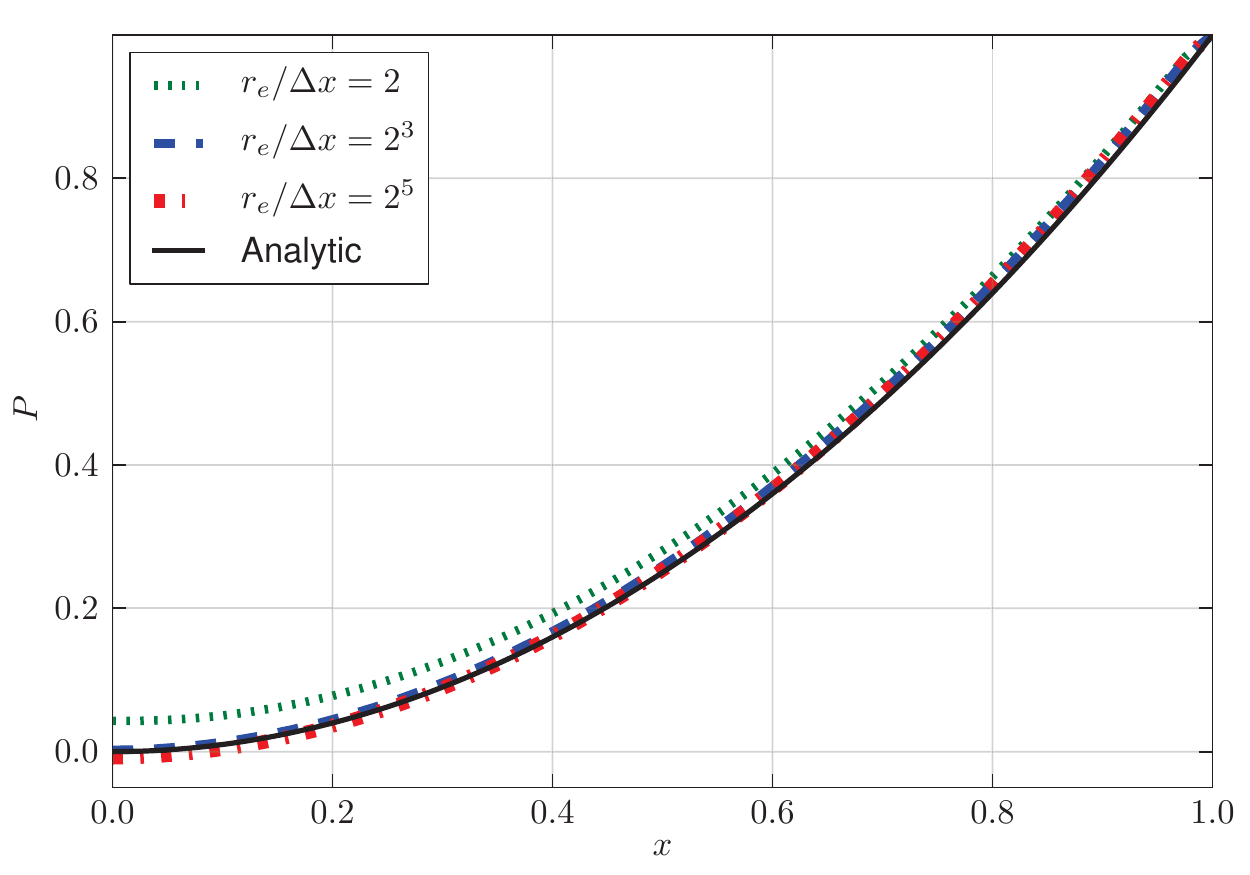}
\caption{Solution of the boundary value problems (\ref{1d_dirichlet}) (left)
and (\ref{1d_neumann}) (right) with corrected MPS symmetrized
Laplacian (\ref{eq:mpslapl_discrete_corrected_symmetrized}).}
\label{fig_poisson1d_corrected_symmetrized}
\end{figure}

The corrected Laplacian formula
(\ref{eq:mpslapl_discrete_corrected}) is now modified
symmetrizing the normalization factors so
that the Poisson equation linear system becomes symmetric.
This is very convenient from a computational point of view.
Additionally this makes the formula conservative in terms
of linear momentum. This is not so relevant in the MPS context
since the gradient formula (the original and the corrected one)
is not intrinsically conservative. Therefore, in the MPS version of
equation (\ref{varmathbbT}) we have a conservative
version of the discrete Laplacian and a non-conservative
one for the pressure gradient. The proposed formula is
\begin{align}
\label{eq:mpslapl_discrete_corrected_symmetrized}
\langle \Delta \phi \rangle_i
&
=
\frac{2d}{\lambda n_0}
\sum_{j\neq i}
\left[
    \frac{1}{\Gamma^\Delta_{ij} }
    \left(
        \phi_j - \phi_i
    \right)
    w
    \left(
            \frac
            {\left|\mathbf{x}_j-\mathbf{x}_i\right|}
            {r_e}
    \right)
\right]
\nonumber
\\
&
+
    \sum_{\mathbf{x}_j\in\partial \Omega}
    \frac{2}{\Gamma^\Delta_{ij} }
    \frac
        {
            \phi_j - \phi_i
        }
        {
            {\left|\mathbf{x}_j-\mathbf{x}_i\right|^2}
        }
        \left(\mathbf{x}_j-\mathbf{x}_i\right)
        W_\Delta\left(\mathbf{x}_j-\mathbf{x}_i ; r_e\right)
        \cdot
        \mathbf{n}_j
        \,
        S_j,
\end{align}
with
\begin{equation}
    \Gamma^\Delta_{ij}
= 0.5\left(\Gamma^\Delta_i + \Gamma^\Delta_j \right).
\end{equation}
As can be seen in figure
\ref{fig_poisson1d_corrected_symmetrized},
the effect of this symmetrization in the
accuracy of the Poisson solver is negligible. For the sake of computational
efficiency this will be the formula used for the free-surface flow discussed in the next section.
%
%
\subsection{Standing Wave}
\label{ss:standing_wave}
The objective of this paper has been to discuss the consistency of the MPS method, to
establish its relationship with SPH,
and to present alternative formulations for some of the operators.
A full assessment of these new formulations when applied to complex 2D or 3D flows
is not within the mentioned objectives and is left for future work. Nonetheless we think it
is interesting to at least present an application of these operators to
a realistic case considering the typical value for the MPS parameters used in
the literature, specially the ratio $r_e/\Delta x$.

The evolution of a viscous standing wave has been chosen.
This is a classical problem in the scientific literature for
which an approximate analytical solution is available \citep{Lighthill2001};
it is of practical interest since it is related to the propagation of gravity waves.
The standing wave flow has been widely studied in SPH (see e.g.
\citet{colagrossi_etal_pre2012_standingwave,colagrossi_etal_pre2011,Antuono2011866,Antuono2012}) and
has been useful as well for validating mesh-based solvers \citep{carrica_etal_ijnmf_2007}.
Periodic boundary conditions have been imposed on the sides of the domain implying
that the surface integrals of the corrected formulations are defined in straight segments
making them easier to compute.

Finally, the particles are in a constant positive pressure state due
to the presence of the hydrostatic pressure which prevents the onset
of tensile instabilities. \citep{Khayyer_gotoh_jcp_2012_mps} performed an interesting
validation study on MPS considering a set of interesting cases (rotating square, impinging jet, etc..)
but not a standing wave one. They were
interested in stability related problems, which is not within our goals.
This is the primary reason for choosing a problem, the standing wave, for
which numerical instabilities are not expected.

\begin{figure}[ht]
\centering
\includegraphics[width=0.695\textwidth]{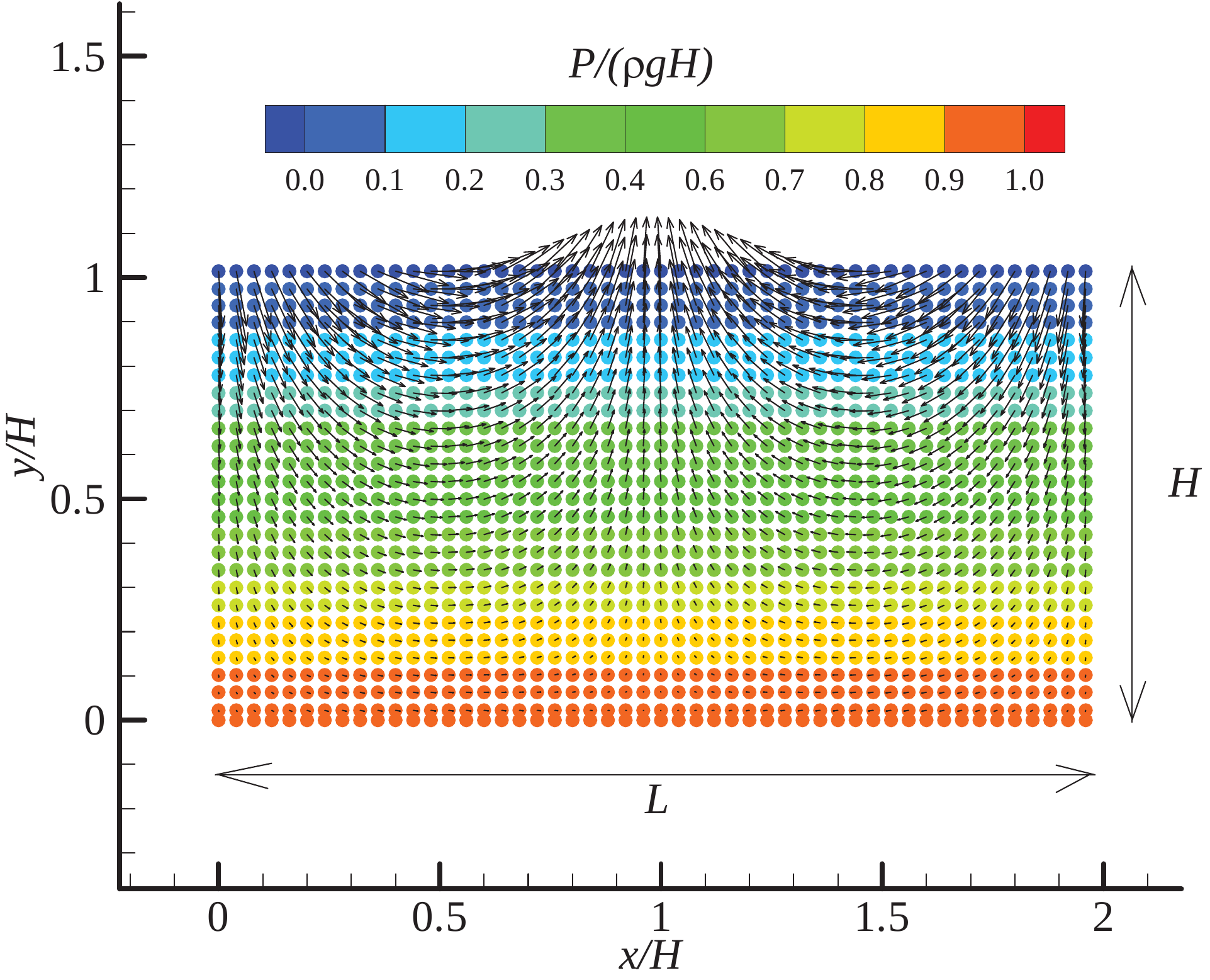}
\caption{Notation and layout of the initial condition for the standing wave problem}
\label{fig_standing_wave}
\end{figure}

The chosen standing wave configuration consists in
a rectangular tank with length $L$ and a water filling height of $H = L/2$.
A sketch of this setup is displayed in figure \ref{fig_standing_wave}.
The wave length is $\lambda = L$,  $k$
is the corresponding wave number ({\em i.e.} $k\,=\,2\pi/\lambda$),
$A$ is the wave amplitude and $\epsilon$ denotes the ratio $2A/H$.
The setup is the same used by \citet{colagrossi_etal_pre2012_standingwave}.

For small-amplitude waves ({\em i.e.} small $\epsilon$),  Potential Theory predicts the following
approximate solution:
\begin{equation}
\begin{array}{ll} \label{pot_standing_wave} \dsty
\varphi(x,y,t)\,=\,\varphi_0(x,y)\,\cos(\omega\,t)\,;\,\quad
\varphi_0(x,y)\,=\,-\,\epsilon\,\frac{H\,g}{2\,\omega}\,
\frac{\cosh[k(y+H)]}{\cosh(kH)}\,\cos(k\,x).
\end{array}
\end{equation}
Here, the circular frequency $\omega$ is given by the dispersion relation of gravity waves, that is,
$\omega^2\,=\,g\,k\,\tanh(k\,H)$ where $g$ is the acceleration of gravity.
At time $t=0$ the free surface is horizontal
while the initial fluid velocity is given by $\nabla \varphi_0$.

%
Potential Theory predicts that the total energy of the standing wave
is conserved during the evolution. However, if the fluid is viscous and the
dissipation due to the solid boundary layers is neglected, it is possible to obtain
an approximate analytical solution that gives the decay of the kinetic energy (see \citep{Lighthill2001}). This is
\begin{equation}\label{Lighthill_diss}
\mathcal{E}_{K}(t)\,=\,\epsilon^2\,g\,\frac{\lambda\,H^2}{32}\,e^{-4\,\nu\,k^2\,t}\,
\big[\,1\,+\,\cos(2\,\omega\,t)\,\big].
\end{equation}
The coefficient of the exponential ($4\,\nu\,k^2$), which governs the kinetic attenuation rate,
depends on the wave number and on the kinematic viscosity $\nu=\mu/\rho$.
The MPS simulations have been implemented by using
a free-slip condition for the velocity
and a Neumann condition for the pressure
along the bottom boundary of the tank, and
periodic BCs on the lateral sides. This is in accordance with the hypothesis made by \citet{Lighthill2001}
in which only internal and free surface viscous boundary layer dissipation sources are considered.

A set of two simulations with parameter values $g=1$, $L=2$, $\epsilon=0.1$ and $H/\Delta x=50, 100$,
using both the classical and corrected formulations have been carried out.
The Reynolds number is $Re =  H\sqrt{gH}/ \nu = 250$, high
enough for the Lighthill approximation to be reasonably accurate.
A value of $r_e/\Delta x=4$ for the Laplacian and $r_e/\Delta x=2$ for the gradient are taken, which
are common for MPS practitioners (see e.g. \citet{youngyoon1999,Tsukamoto_cheng_2011}).

\begin{figure}[ht]
\centering
\includegraphics[width=0.95\textwidth]{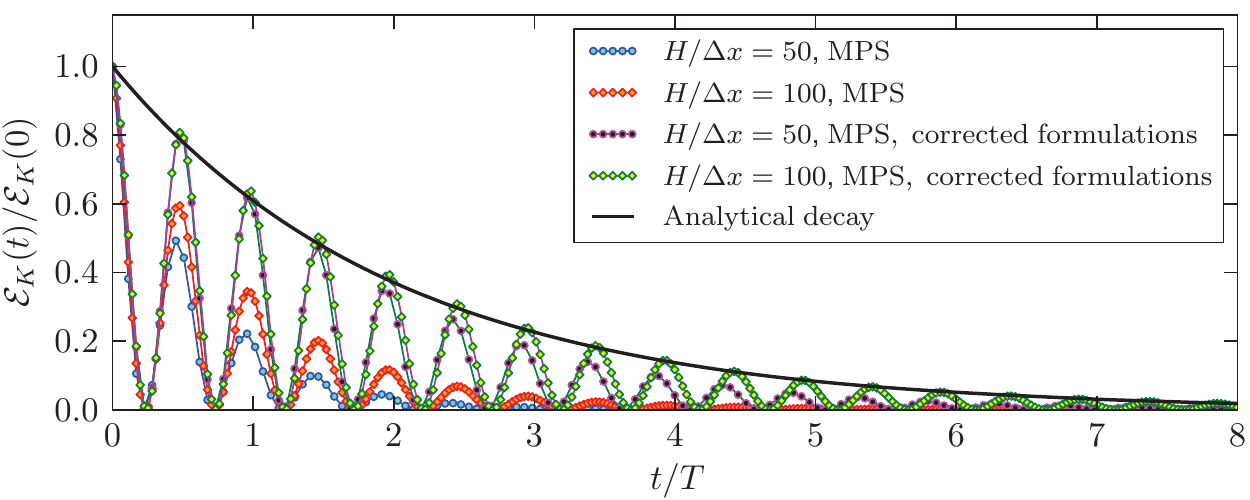}
\caption{Standing wave kinetic energy time evolution }
\label{fig_provisionalRe250}
\end{figure}

\begin{figure}[ht]
\centering
\includegraphics[width=0.495\textwidth]{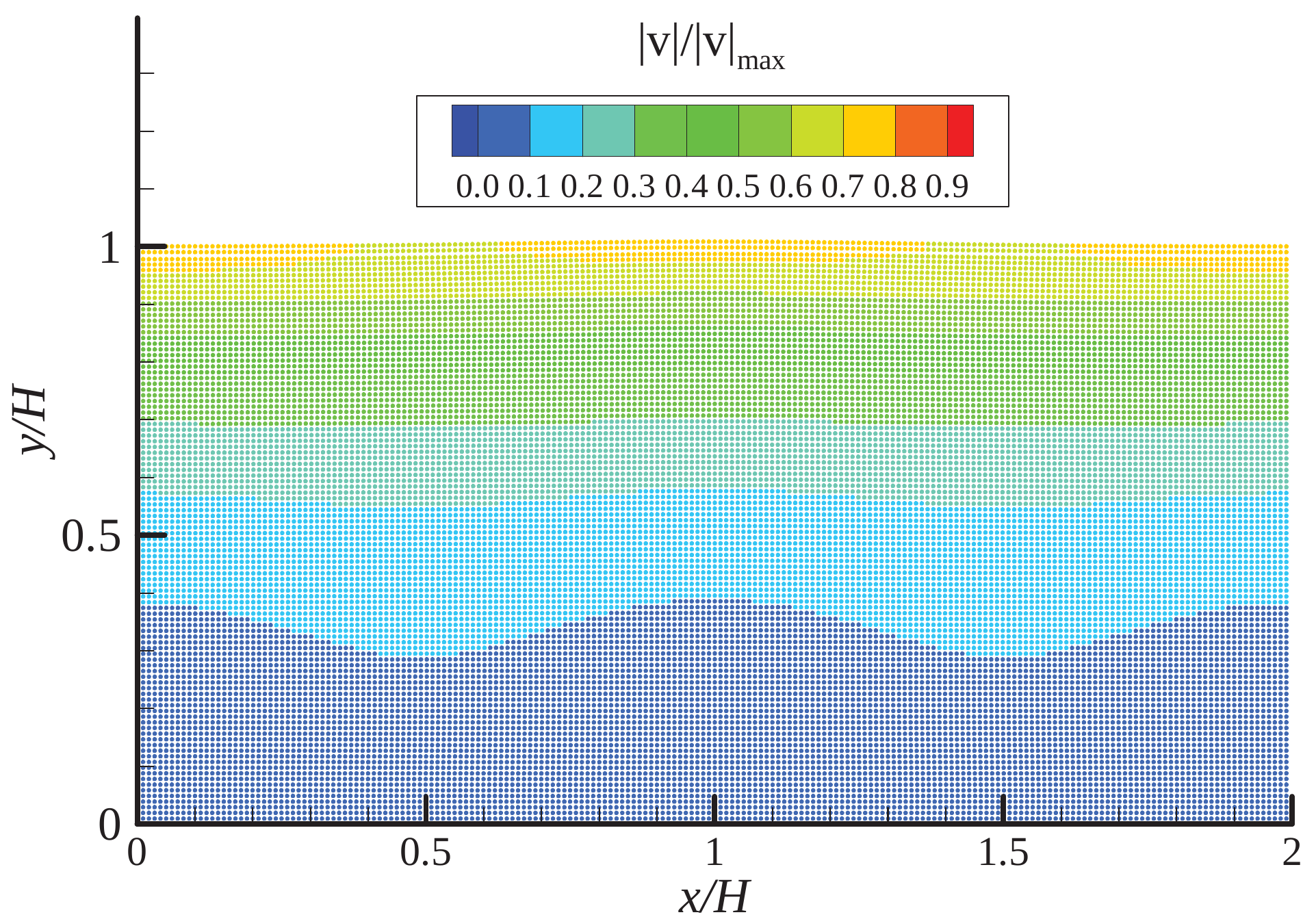}
\includegraphics[width=0.495\textwidth]{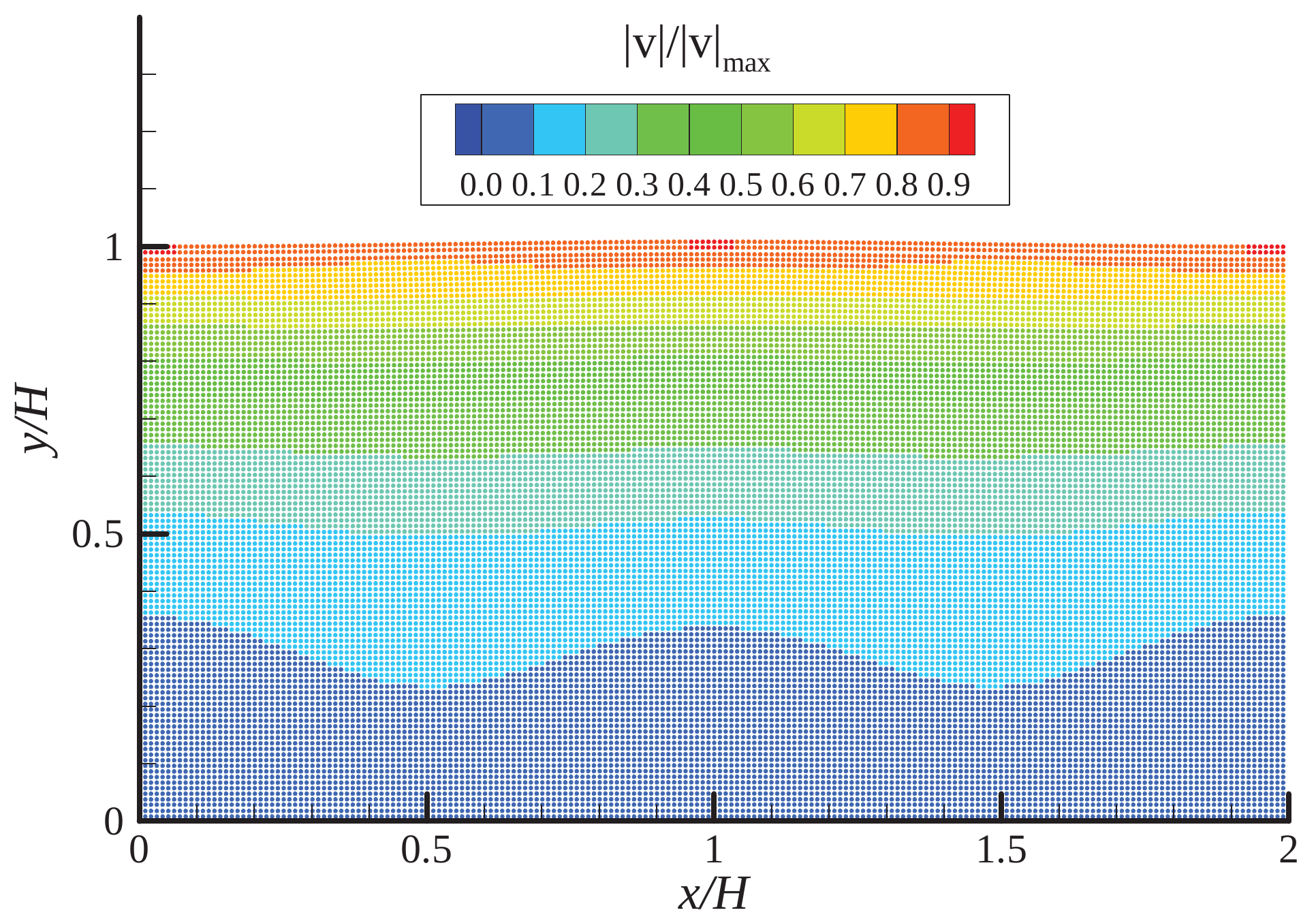}
\caption{Velocity field modulus at the second kinetic energy peak, $t/T\approx 0.5$, uncorrected formulation (left), corrected (right).}
\label{fig_standing_wave2}
\end{figure}

In figure \ref{fig_provisionalRe250} the decay of the kinetic energy is displayed using
a classical MPS scheme and the proposed corrected formulas discussed in section \ref{ss:consistent_def_mps_operators}.
The corrected approach performs better when compared to the analytical solution with
significant over-damping observed in the classical approach. A plot of the velocity field modulus after half an oscillation cycle,
normalized with the maximum initial velocity,
is provided in figure \ref{fig_standing_wave2}. The
graphs show the extra damping occurring in the classical approach simulations, made
noticeable by the differences in the velocity modulus in the area close to the free surface.
This evidence is not enough to proclaim any general conclusion although the results are promising.
Further investigation is left for future work.
%
\section{Conclusions}
The consistency of the moving particle semi-implicit (MPS) method in
reproducing the gradient, divergence and Laplacian
differential operators has been discussed in the present paper.
It has been shown that the operators are consistent by performing
an analogy with the SPH formulations. The equivalence between the
MPS weighting function and the SPH kernel has been established. It has
been shown that this equivalence differs for the first order and
second order differential operators leading to different equivalent
SPH kernels formulations. It has been shown that for each MPS weighting
function it is possible to find two SPH kernels, one which allows
the computation of the MPS gradient and a different one for the Laplacian.
It has been shown that the converse also holds;
from an SPH kernel two MPS weighting
functions can be obtained, one using the gradient equivalence and another,
the Laplacian equivalence. To summarize, we show that SPH and MPS are closely related.

SPH started earlier in the seventies and was applied in the early nineties
to free-surface flows using an explicit approach with a weakly compressible
fluid model to numerically simulate liquid behavior.
Later in the mid nineties, the MPS method appeared
imposing incompressibility with a projection scheme.
In the late nineties a similar approach was followed
to obtain the first incompressible SPH model.
From then on, they have run parallel, approaching a wide range of problems.
The connections amongst them have been clearly established in the
present paper through the comparison of the differential operators
and the convolution functions.
We believe this can be useful in order to establish a common framework
that can help the progress of both methods.

The application of the MPS scheme in solving the Navier-Stokes equations, using
a fractional step approach, has been treated.
Some inconsistency problems when solving the Poisson equation in the
fractional step method have been unveiled.
It has been shown that in general, such an equation does
not have a solution when using the MPS integral formulation.
Due to the equivalences shown, the same
inconsistency problems occur with the incompressible SPH method.

A corrected formulation of the operators has been proposed involving
the use of boundary integrals. Applications to one dimensional boundary value
Dirichlet and mixed Neumann-Dirichlet problems have been presented.
These applications show that the use of the corrected formulations is
crucial in order to obtain a convergence to the exact results in all these
problems.

Although in this work we have not focused on validation of practical
fluid mechanics problems, a complex free-surface flow application, namely 
the evolution of a viscous standing wave
has been briefly discussed. We have shown that the corrected formulations improve
the accuracy of the results under the analyzed conditions.

As a final remark, and after
showing the equivalence of MPS and SPH methods,
we would like to stress that, although being crucial in guaranteeing
convergence to the exact solution,
consistency is not the only factor to take into account in order
to have a reliable numerical method.
Stability, conservation properties and computational
feasibility are equally important.
Translating all the existing results available in the MPS literature
to SPH and viceversa is left, amongst other things, as future work.
\begin{ack}
The research leading to these results has received funding from
the Spanish Ministry for Science and Innovation under
grant TRA2010-16988 ``\textit{Ca\-rac\-te\-ri\-za\-ci\'on Num\'erica y
Experimental de las Cargas Fluido-Din\'amicas en el transporte de Gas Licuado}'' .

The authors thank Juan J. L. Vel\'{a}zquez for the valuable conversations
regarding the results presented in section \ref{s:bc}.
The authors are grateful to Hugo Gee for the correction and
improvement of the English text. Finally,
the authors are particularly grateful to Andrea Colagrossi
from INSEAN-Rome; collaborating with him during the recent years has been crucial for
the progress of our research group.
\end{ack}


\begin{thebibliography}{48}
\expandafter\ifx\csname natexlab\endcsname\relax\def\natexlab#1{#1}\fi
\expandafter\ifx\csname url\endcsname\relax
  \def\url#1{\texttt{#1}}\fi
\expandafter\ifx\csname urlprefix\endcsname\relax\def\urlprefix{URL }\fi

\bibitem[{Amicarelli et~al.(2011)Amicarelli, Marongiu, Leboeuf, Leduc, and
  Caro}]{Amicarelli2011279}
Amicarelli, A., Marongiu, J.-C., Leboeuf, F., Leduc, J., Caro, J., 2011. {SPH}
  truncation error in estimating a {3D} function. Computers \& Fluids 44~(1),
  279 -- 296.
\newline\urlprefix\url{http://www.sciencedirect.com/science/article/B6V26-5207%
9TP-1/2/28aecd9efa85d26137ba55bcaa45daa4}

\bibitem[{Antuono and Colagrossi(2012)}]{Antuono2012}
Antuono, M., Colagrossi, A., 2012. The damping of viscous gravity waves. Wave
  Motion~(0), --.
\newline\urlprefix\url{http://www.sciencedirect.com/science/article/pii/S01652%
12512001060}

\bibitem[{Antuono et~al.(2011)Antuono, Colagrossi, Marrone, and
  Lugni}]{Antuono2011866}
Antuono, M., Colagrossi, A., Marrone, S., Lugni, C., 2011. Propagation of
  gravity waves through an {SPH} scheme with numerical diffusive terms.
  Computer Physics Communications 182~(4), 866 -- 877.
\newline\urlprefix\url{http://www.sciencedirect.com/science/article/B6TJ5-51PR%
YCS-1/2/b69f4a14370f08b80c197c90035b226a}

\bibitem[{Benz(1988)}]{benz1988}
Benz, W., 1988. Applications of smoothed particle hydrodynamics (sph) to
  astrophysical problems. Computer Physics Communications 48, 97--105.

\bibitem[{Carrica et~al.(2007)Carrica, Wilson, and
  Stern}]{carrica_etal_ijnmf_2007}
Carrica, P.~M., Wilson, R.~V., Stern, F., 2007. An unsteady single-phase level
  set method for viscous free surface flows. International Journal for
  Numerical Methods in Fluids 53~(2), 229--256.
\newline\urlprefix\url{http://dx.doi.org/10.1002/fld.1279}

\bibitem[{Chorin(1968)}]{chorin_fractional_step_poisson_1968}
Chorin, A.~J., Oct. 1968. {Numerical Solution of the Navier-Stokes Equations}.
  Mathematics of Computation 22~(104), 745--762.
\newline\urlprefix\url{http://dx.doi.org/10.2307/2004575}

\bibitem[{Colagrossi et~al.(2011)Colagrossi, Antuono, Souto-Iglesias, and
  Le~Touz\'{e}}]{colagrossi_etal_pre2011}
Colagrossi, A., Antuono, M., Souto-Iglesias, A., Le~Touz\'{e}, D., 2011.
  {Theoretical analysis and numerical verification of the consistency of
  viscous smoothed-particle-hydrodynamics formulations in simulating
  free-surface flows}. Physical Review E 84, 26705+.

\bibitem[{Colagrossi et~al.(2009)Colagrossi, Antuono, and
  Touz\'{e}}]{Colagrossi2009}
Colagrossi, A., Antuono, M., Touz\'{e}, D.~L., 2009. Theoretical considerations
  on the free-surface role in the {S}moothed-particle-hydrodynamics model.
  Physical Review E (Statistical, Nonlinear, and Soft Matter Physics) 79~(5),
  056701.

\bibitem[{Colagrossi et~al.(2012)Colagrossi, Souto-Iglesias, Antuono, and
  Marrone}]{colagrossi_etal_pre2012_standingwave}
Colagrossi, A., Souto-Iglesias, A., Antuono, M., Marrone, S., 2012. {SPH}
  modeling of dissipation mechanisms in gravity waves. (submitted for
  publication).

\bibitem[{Cummins and Rudman(1999)}]{cummins1999}
Cummins, S., Rudman, M., July 1999. An {SPH} projection method. J. Comp. Phys.
  152~(2), 584--607.

\bibitem[{Dalrymple and Rogers(2006)}]{dalrymple2006}
Dalrymple, R., Rogers, B., 2006. Numerical modeling of water waves with the sph
  method. Coastal Engineering 53 (2-3), 141--147.

\bibitem[{De~Leffe et~al.(2009)De~Leffe, Le~Touz\'{e}, and
  Alessandrini}]{deleffe_etal_spheric09}
De~Leffe, M., Le~Touz\'{e}, D., Alessandrini, B., May 2009. Normal flux method
  at the boundary for {SPH}. In: $4^{th}$ SPHERIC. pp. 149--156.

\bibitem[{Dehnen and Aly(2012)}]{dehnen_aly_wendland_2012}
Dehnen, W., Aly, H., 2012. Improving convergence in smoothed particle
  hydrodynamics simulations without pairing instability. Monthly Notices of the
  Royal Astronomical Society 425~(2), 1068--1082.
\newline\urlprefix\url{http://dx.doi.org/10.1111/j.1365-2966.2012.21439.x}

\bibitem[{Espa\~nol and Revenga(2003)}]{espanol2003}
Espa\~nol, P., Revenga, M., Feb 2003. Smoothed dissipative particle dynamics.
  Phys. Rev. E 67~(2), 026705.

\bibitem[{Ferrand et~al.(2012)Ferrand, Laurence, Rogers, Violeau, and
  Kassiotis}]{ferrand_etal_2012}
Ferrand, M., Laurence, D., Rogers, B., Violeau, D., Kassiotis, C., 2012.
  Unified semi-analytical wall boundary conditions for inviscid, laminar or
  turbulent flows in the meshless {SPH} method. International Journal for
  Numerical Methods in Fluids, n/a--n/a.
\newline\urlprefix\url{http://dx.doi.org/10.1002/fld.3666}

\bibitem[{Gomez-Gesteira et~al.(2010)Gomez-Gesteira, Rogers, Dalrymple, and
  Crespo}]{gomezgesteira_etal_jhr2010}
Gomez-Gesteira, M., Rogers, B.~D., Dalrymple, R.~A., Crespo, A. J.~C., 2010.
  {State-of-the-art of classical SPH for free-surface flows}. Journal of
  Hydraulic Research 48~(S1), 6--27.
\newline\urlprefix\url{http://dx.doi.org/10.1080/00221686.2010.9641242}

\bibitem[{Khayyer and Gotoh(2009{\natexlab{a}})}]{Khayyer_gotoh_ce09}
Khayyer, A., Gotoh, H., 2009{\natexlab{a}}. Modified moving particle
  semi-implicit methods for the prediction of 2d wave impact pressure. Coastal
  Engineering 56~(4), 419 -- 440.
\newline\urlprefix\url{http://www.sciencedirect.com/science/article/pii/S03783%
83908001646}

\bibitem[{Khayyer and Gotoh(2009{\natexlab{b}})}]{Khayyer_gotoh_ijope09}
Khayyer, A., Gotoh, H., December 2009{\natexlab{b}}. Wave impact pressure
  calculations by improved {SPH} methods. International Journal of Offshore and
  Polar Engineering 19~(4), 300--307.

\bibitem[{Khayyer and Gotoh(2010)}]{Khayyer2010_aor}
Khayyer, A., Gotoh, H., 2010. A higher order laplacian model for enhancement
  and stabilization of pressure calculation by the mps method. Applied Ocean
  Research 32~(1), 124 -- 131.
\newline\urlprefix\url{http://www.sciencedirect.com/science/article/pii/S01411%
18710000027}

\bibitem[{Khayyer and Gotoh(2011)}]{Khayyer_gotoh_jcp_2012_mps}
Khayyer, A., Gotoh, H., 2011. Enhancement of stability and accuracy of the
  moving particle semi-implicit method. Journal of Computational Physics
  230~(8), 3093 -- 3118.
\newline\urlprefix\url{http://www.sciencedirect.com/science/article/pii/S00219%
99111000271}

\bibitem[{Khayyer and Gotoh(2012)}]{Khayyer2012_aor}
Khayyer, A., Gotoh, H., 2012. A 3d higher order laplacian model for enhancement
  and stabilization of pressure calculation in 3d mps-based simulations.
  Applied Ocean Research 37~(0), 120 -- 126.
\newline\urlprefix\url{http://www.sciencedirect.com/science/article/pii/S01411%
18712000399}

\bibitem[{Koshizuka et~al.(1998)Koshizuka, Nobe, and Oka}]{koshizuka1998}
Koshizuka, S., Nobe, A., Oka, Y., 1998. Numerical analysis of breaking waves
  using the moving particle semi-implicit method. International Journal for
  Numerical Methods in Fluids 26~(7), 751--769.

\bibitem[{Koshizuka and Oka(1996)}]{koshizuka1996}
Koshizuka, S., Oka, Y., 1996. Moving-particle semi-implicit method for
  fragmentation of incompressible fluid. Nuclear Science and Engineering
  123~(3), 421--434.

\bibitem[{Koshizuka et~al.(1995)Koshizuka, Oka, and Tamako}]{koshizuza1995}
Koshizuka, S., Oka, Y., Tamako, H., 1995. A particle method for calculating
  splashing of incompressible viscous fluid. In: International Conference,
  Mathematics and Computations, Reactor Physics, and Environmental Analyses.
  Vol.~2. pp. 1514--1521.

\bibitem[{{Laibe} and {Price}(2012{\natexlab{a}})}]{laibe_price_2012a}
{Laibe}, G., {Price}, D.~J., Mar. 2012{\natexlab{a}}. {Dusty gas with smoothed
  particle hydrodynamics - I. Algorithm and test suite}. Monthly Notices of the
  RAS 420, 2345--2364.

\bibitem[{{Laibe} and {Price}(2012{\natexlab{b}})}]{laibe_price_2012b}
{Laibe}, G., {Price}, D.~J., Mar. 2012{\natexlab{b}}. {Dusty gas with smoothed
  particle hydrodynamics - II. Implicit timestepping and astrophysical drag
  regimes}. Monthly Notices of the RAS 420, 2365--2376.

\bibitem[{Lee et~al.(2008)Lee, Moulinec, Xu, Violeau, Laurence, and
  Stansby}]{lee2008}
Lee, E.~S., Moulinec, C., Xu, R., Violeau, D., Laurence, D., Stansby, P., 9/10
  2008. Comparisons of weakly compressible and truly incompressible algorithms
  for the {SPH} mesh free particle method. Journal of Computational Physics
  227~(18), 8417--8436.

\bibitem[{Libersky et~al.(1993)Libersky, Petschek, Carney, Hipp, and
  Allahdadi}]{libersky1993}
Libersky, L., Petschek, A., Carney, T., Hipp, J., Allahdadi, F., 1993. High
  strain lagrangian hydrodynamics a three-dimensional sph code for dynamic
  material response. J. Comp. Phys. 109~(1), 67--75.

\bibitem[{Lighthill(2001)}]{Lighthill2001}
Lighthill, J., 2001. Waves in fluids. Cambridge University Press.

\bibitem[{Maci\`{a} et~al.(2011{\natexlab{a}})Maci\`{a}, Antuono, Gonz\'{a}lez,
  and Colagrossi}]{MaciaetalPTP}
Maci\`{a}, F., Antuono, M., Gonz\'{a}lez, L.~M., Colagrossi, A.,
  2011{\natexlab{a}}. Theoretical analysis of the no-slip boundary condition
  enforcement in {SPH} methods. Progress of Theoretical Physics 125~(6),
  1091--1121.
\newline\urlprefix\url{http://ptp.ipap.jp/link?PTP/125/1091/}

\bibitem[{Maci\`{a} et~al.(2011{\natexlab{b}})Maci\`{a}, Colagrossi, Antuono,
  and Souto-Iglesias}]{macia_etal_spheric11_wendland}
Maci\`{a}, F., Colagrossi, A., Antuono, M., Souto-Iglesias, A.,
  2011{\natexlab{b}}. {Benefits of using a Wendland kernel for free-surface
  flows}. In: 6th ERCOFTAC SPHERIC workshop on SPH applications. Hamburg
  University of Technology, pp. 30--37.

\bibitem[{Maci\`{a} et~al.(2012)Maci\`{a}, Gonz\'{a}lez, Cercos-Pita, and
  Souto-Iglesias}]{Maciaetal_PTP_2012}
Maci\`{a}, F., Gonz\'{a}lez, L.~M., Cercos-Pita, J.~L., Souto-Iglesias, A.,
  Sep. 2012. A boundary integral {SPH} formulation. {C}onsistency and
  applications to {ISPH} and {WCSPH}. Progress of Theoretical Physics 128~(3).
\newline\urlprefix\url{\#}

\bibitem[{Monaghan(1994)}]{mon1994b}
Monaghan, J., 1994. Simulating free surface flows with {SPH}. J. Comp. Phys.
  110~(2), 39--406.

\bibitem[{Monaghan(2012)}]{monaghan_arfm_2012}
Monaghan, J., 2012. Smoothed particle hydrodynamics and its diverse
  applications. Annual Review of Fluid Mechanics 44~(1), 323--346.
\newline\urlprefix\url{http://www.annualreviews.org/doi/abs/10.1146/annurev-fl%
uid-120710-101220}

\bibitem[{Monaghan(2005)}]{mon2005}
Monaghan, J.~J., 2005. Smoothed particle hydrodynamics. Reports on Progress in
  Physics 68, 1703--1759.

\bibitem[{Morris et~al.(1997)Morris, Fox, and Zhu}]{Morris+etal:1997}
Morris, J.~P., Fox, P.~J., Zhu, Y., 1997. Modeling low {R}eynolds number
  incompressible flows using {SPH}. Journal of Computational Physics 136,
  214--226.

\bibitem[{Naito and Sueyoshi(2001)}]{naito2001}
Naito, S., Sueyoshi, M., Sept. 2001. A numerical analysis of violent free
  surface flow on flooded car deck using particle method. In: 5th International
  Workshop Stability and Operational Safety of Ships. Trieste.

\bibitem[{Price(2012)}]{Price2012_jcp}
Price, D.~J., 2012. Smoothed particle hydrodynamics and magnetohydrodynamics.
  Journal of Computational Physics 231~(3), 759 -- 794.
\newline\urlprefix\url{http://www.sciencedirect.com/science/article/pii/S00219%
99110006753}

\bibitem[{Quinlan et~al.(2006)Quinlan, Lastiwka, and Basa}]{Quinlan_06}
Quinlan, N.~J., Lastiwka, M., Basa, M., 2006. Truncation error in mesh-free
  particle methods. International Journal for Numerical Methods in Engineering
  66~(13), 2064--2085.
\newline\urlprefix\url{http://dx.doi.org/10.1002/nme.1617}

\bibitem[{Robinson(2009)}]{Robinson_PhDthesis}
Robinson, M., 2009. {Turbulence and Viscous Mixing using Smoothed Particle
  Hydrodynamics}. Ph.D. thesis, Department of Mathematical Science, Monash
  University.

\bibitem[{Sueyoshi and Naito(2003)}]{sueyoshi2003}
Sueyoshi, M., Naito, S., Sept. 2003. A numerical study of violent free surface
  problems with particle method for marine engineering. In: 8th International
  Conference on Numerical Ship Hydrodynamics, Busan(Korea). pp. 330--339.

\bibitem[{Tanaka and Masunaga(2010)}]{Tanaka_etal_jcp2010_mps}
Tanaka, M., Masunaga, T., 2010. Stabilization and smoothing of pressure in mps
  method by quasi-compressibility. Journal of Computational Physics 229~(11),
  4279 -- 4290.
\newline\urlprefix\url{http://www.sciencedirect.com/science/article/pii/S00219%
99110000847}

\bibitem[{Tsukamoto et~al.(2011)Tsukamoto, Cheng, and
  Nishimoto}]{Tsukamoto_cheng_2011}
Tsukamoto, M.~M., Cheng, L.-Y., Nishimoto, K., 2011. Analytical and numerical
  study of the effects of an elastically-linked body on sloshing. Computers \&
  Fluids 49~(1), 1 -- 21.
\newline\urlprefix\url{http://www.sciencedirect.com/science/article/pii/S00457%
93011001423}

\bibitem[{Valizadeh and Monaghan(2012)}]{valizadeh_monaghan_pof2012}
Valizadeh, A., Monaghan, J.~J., 2012. Smoothed particle hydrodynamics
  simulations of turbulence in fixed and rotating boxes in two dimensions with
  no-slip boundaries. Physics of Fluids 24~(3), 035107.
\newline\urlprefix\url{http://link.aip.org/link/?PHF/24/035107/1}

\bibitem[{Violeau(2009)}]{violeau_2009}
Violeau, D., Sep 2009. Dissipative forces for lagrangian models in
  computational fluid dynamics and application to smoothed-particle
  hydrodynamics. Phys. Rev. E 80, 036705.
\newline\urlprefix\url{http://link.aps.org/doi/10.1103/PhysRevE.80.036705}

\bibitem[{Violeau(2012)}]{violeau_2012_book}
Violeau, D., 2012. {Fluid Mechanics and the SPH Method}. Oxford University
  Press.

\bibitem[{Yoon et~al.(1999{\natexlab{a}})Yoon, Koshizuka, and
  Oka}]{youngyoon1999b}
Yoon, H.-Y., Koshizuka, S., Oka, Y., 1999{\natexlab{a}}. A mesh-free numerical
  method for direct simulation of gas-liquid phase interface.
  Nuclear-Science-and-Engineering 133~(2), 192--200.

\bibitem[{Yoon et~al.(1999{\natexlab{b}})Yoon, Koshizuka, and
  Oka}]{youngyoon1999}
Yoon, H.-Y., Koshizuka, S., Oka, Y., 1999{\natexlab{b}}. A particle-gridless
  hybrid method for incompressible flows. International Journal for Numerical
  Methods in Fluids 30~(4), 407--424.

\end{thebibliography}

\appendix
\section{SPH approximation of differential operators}
\subsection{General}
Smoothed Particle Hydrodynamics (SPH) is a numerical method used among many other applications
to solve the Navier-Stokes equations. It has very attractive features for certain types of flows, namely that
it is mesh-free, its Lagrangian character and its conservation properties.
SPH has been used in a wide range of contexts, including Astrophysics \citep{benz1988},
Magnetohydrodynamics \citep{Price2012_jcp}, free surface flows \citep{mon1994b,gomezgesteira_etal_jhr2010}, Coastal
Engineering applications \citep{dalrymple2006} and even Solid Mechanics \citep{libersky1993}.
A comprehensive recent review of the method can be found in \citep{violeau_2012_book}.
In SPH both a weakly compressible approach in order to impose incompressibility
\citep{mon2005} and a pure incompressible one \citep{cummins1999,lee2008} similar
to MPS have coexisted through the years.

In this appendix, the continuous version of the SPH
modeling of the gradient and Laplacian differential operators together with
some related consistency statements are introduced;
they are useful in describing the equivalences
with the MPS method discussed in the bulk of the paper.
\subsection{Kernel}
Let $W\left(  \mathbf{x};h\right)  $ be the function of $\mathbf{x}\in \mathbb{R}^d$ depending on $h>0$
defined by
\begin{equation}
\label{eq:sphkernel}
W\left(  \mathbf{x};h\right)  =\frac{1}{h^{d}}\tilde{W}\left(  \left\vert \frac{\mathbf{x}}{h}\right\vert \right)  \text{,}%
\end{equation}
where $\tilde{W}:\mathbb{R\rightarrow R}$ is a \emph{nonnegative
differentiable function} such that:%
\begin{equation}
\int_{\mathbb{R}^{d}}
\tilde{W}\left(
\left\vert \mathbf{x}\right\vert
\right)
d\mathbf{x}=1.
\end{equation}
When $d>1$
\begin{equation}
\int_{\mathbb{R}^{d}}\tilde{W}\left(  \left\vert \mathbf{x}\right\vert
\right)  d\mathbf{x}=\omega_{d}\int_{0}^{\infty}\tilde{W}\left(  r\right)
r^{d-1}dr,
\end{equation}
the constant $\omega_{d}$ being the volume of the unit sphere in
$\mathbb{R}^{d}$.

An important result used in the calculations
documented in the paper in order to establish the relationships existing
between the SPH kernel and the MPS weighting functions is the following:
\begin{equation}
\label{eq:wgradw_d}
\int_{\mathbb{R}^d}
    \tilde{W}(\left|\mathbf{x}\right|) d\mathbf{x}
=
-
\frac{1}{d}
\int_{\mathbb{R}^d}
\left|\mathbf{x}\right|
\tilde{W}^{\prime}(\left|\mathbf{x}\right|)d\mathbf{x}.
\end{equation}
This relationship is a direct consequence of the following identities
obtained through integration by parts
\begin{displaymath}
-
\int_{\mathbb{R}^d}
    \frac{x_k^2}{\left|\mathbf{x}\right|}
\tilde{W}^{\prime}(\left|\mathbf{x}\right|)d\mathbf{x}
=
-\int_{\mathbb{R}^d}
    x_k
    \partial_{x_k}
    \tilde{W}(\left|\mathbf{x}\right|)d\mathbf{x}
=
    \int_{\mathbb{R}^d}
    \tilde{W}(\left|\mathbf{x}\right|) d\mathbf{x},
\end{displaymath}
after summing over $k$.
\subsection{Continuous SPH approximation}
The SPH approximation with respect to the kernel $W$ of a function $u\left(
\mathbf{x}\right)  $ on $\mathbb{R}^{d}$ taking real values is defined as:%
\begin{equation}
\left\langle u\right\rangle \left(  \mathbf{x}\right)  :=%
{\displaystyle\int_{\mathbb{R}^{d}}}
u\left(  \mathbf{x}^{\prime}\right)  W\left(  \mathbf{x-x}^{\prime};h\right)
d\mathbf{x}^{\prime}%
\end{equation}
\subsection{Derivatives\textbf{ }}
 \label{ss:appendix:grad}
The SPH approximation of the partial derivative $\partial_{x_{k}}u$ is set to
be:%
\begin{equation}
\left\langle \partial_{x_{k}}u\right\rangle \left(  \mathbf{x}\right)
:=\int_{\mathbb{R}^{d}}u\left(  \mathbf{x}^{\prime}\right)  \partial_{x_{k}%
}W\left(  \mathbf{x-x}^{\prime};h\right)  d\mathbf{x}^{\prime}.
\end{equation}
Note that $\left\langle \partial_{x_{k}}u\right\rangle $ coincides with
$\partial_{x_{k}}\left\langle u\right\rangle $.
In addition, since
\[
\nabla_{\mathbf{x}}W\left(  \mathbf{x;}h\right)  =\frac{1}{h^{d+1}}\tilde
{W}^{\prime}\left(  \frac{\left\vert \mathbf{\mathbf{x}}\right\vert }%
{h}\right)  \frac{\mathbf{x}}{\left\vert \mathbf{x}\right\vert },
\]
we can write%
\begin{equation}
\label{eq:gradsphcont}
\left\langle \partial_{x_{k}}u\right\rangle \left(  \mathbf{x}\right)
=\frac{1}{h^{d+1}}\int_{\mathbb{R}^{d}}u\left(  \mathbf{x}^{\prime}\right)
\frac{x_{k}-x_{k}^{^{\prime}}}{\left\vert \mathbf{x}-\mathbf{x}^{\prime
}\right\vert }\tilde{W}^{\prime}\left(  \frac{\left\vert \mathbf{\mathbf{x-x}%
}^{\prime}\right\vert }{h}\right)  d\mathbf{x}^{\prime}
\end{equation}
The previous expression can be anti-symmetrized so that, at the discrete level,
the derivative of a constant function is identically zero. Note that:
\begin{displaymath}
\int_{\mathbb{R}^{d}}
\frac
{
    {x_{k}-x_{k}^{\prime}}
}
{
    \left\vert \mathbf{x}-\mathbf{x}^{\prime}\right\vert
}
\tilde{W}^{\prime}
\left(  \frac{\left\vert \mathbf{\mathbf{x-x}%
}^{\prime}\right\vert }{h}\right)
d\mathbf{x}^{\prime}=0.
\end{displaymath}
Therefore (\ref{eq:gradsphcont}) is also equal to:
\begin{equation}
\label{eq:contsphgrad}
\left\langle \partial_{x_{k}}u\right\rangle
\left(
\mathbf{x}
\right)
=
\frac{1}{h^{d+1}}\int_{\mathbb{R}^{d}}
\frac{
    u
    \left(  \mathbf{x}^{\prime}\right)
    -
    u
    \left(  \mathbf{x}\right)
}
{
    \left\vert \mathbf{x}-\mathbf{x}^{\prime}\right\vert
}
\left({x_{k}-x_{k}^{\prime}}\right)
\tilde{W}^{\prime}
\left(  \frac{\left\vert \mathbf{\mathbf{x-x}%
}^{\prime}\right\vert }{h}\right)
d\mathbf{x}^{\prime}.
\end{equation}
Nonetheless, the discrete version of this formula does not conserve 
linear momentum when used to evaluate the pressure gradient 
and it is generally replaced by a similar one
which is instead symmetric (see \citep{mon2005} for a specific
discussion on this topic).

%
The divergence operator can be written as:
\begin{equation}
\label{eq:contsphdiv}
\left\langle \nabla \cdot \mathbf{u} \right\rangle
\left(
\mathbf{x}
\right)
=
\frac{1}{h^{d+1}}\int_{\mathbb{R}^{d}}
\frac{
        \left(
            \mathbf{u}
            \left(
                \mathbf{x}^{\prime}
            \right) 
            - 
            \mathbf{u}
            \left(
                \mathbf{x}
            \right)
        \right)
        \cdot
        \left(
            \mathbf{x} - \mathbf{x^\prime}
        \right)
}
{
    \left\vert \mathbf{x}-\mathbf{x}^{\prime}\right\vert
}
\tilde{W}^{\prime}
\left(  \frac{\left\vert \mathbf{\mathbf{x-x}%
}^{\prime}\right\vert }{h}\right)
d\mathbf{x}^{\prime},
\end{equation}
which is identically zero for constant fields at the discrete level.
\subsection{The Laplacian}
\label{ss:app:lapl:morris}
We will consider the following SPH approximation of the Laplacian of a
function $u\left(  \mathbf{x}\right)  $ due to \citet{Morris+etal:1997}
whose consistency was demonstrated by \citet{espanol2003}.
\begin{equation}
\label{eq:a11}
\left\langle \Delta u\right\rangle\left(  \mathbf{x}\right)
:=2\int_{\mathbb{R}^{d}}\frac{\left(  \mathbf{x}^{\prime}-\mathbf{x}\right)
\cdot\nabla_{\mathbf{x}}W\left(  \mathbf{x}^{\prime}-\mathbf{x;}h\right)
}{\left\vert \mathbf{x}^{\prime}-\mathbf{x}\right\vert ^{2}}\left[  u\left(
\mathbf{x}^{\prime}\right)  -u\left(  \mathbf{x}\right)  \right]
d\mathbf{x}^{\prime}.
\end{equation}
In order to compare this expression to the MPS approximation of the Laplacian, it
is useful to rewrite (\ref{eq:a11})as: 
\begin{equation}
\label{eq:sphlaplmorris}
\left\langle \Delta u\right\rangle\left(  \mathbf{x}\right)
=-\frac{2}{h^{d+1}}\int_{\mathbb{R}^{d}}\frac{u\left(  \mathbf{x}^{\prime
}\right)  -u\left(  \mathbf{x}\right)  }{\left\vert \mathbf{x}^{\prime
}-\mathbf{x}\right\vert }\tilde{W}^{\prime}\left(  \frac{\left\vert
\mathbf{\mathbf{x}^{\prime}-\mathbf{x}}\right\vert }{h}\right)  d\mathbf{x}%
^{\prime}.
\end{equation}
\subsection{Consistency}
\label{ss:app:sphconsistency}
The reader is referred to \citep{MaciaetalPTP} for details on consistency results
of SPH approximation to differential operators. A summary is provided here.
For a general twice differentiable function $u$ the following holds:%
\[
\left\langle u\right\rangle =u+\mathcal{O}\left(  h^{2}\right)  ,
\]%
\[
\left\langle \nabla_{\mathbf{x}}u\right\rangle =\nabla_{\mathbf{x}}u +\mathcal{O}\left(h^{2}\right),
\qquad
\langle \nabla \cdot \mathbf{u} \rangle = \nabla \cdot \mathbf{u}+\mathcal{O}\left(h^{2}\right),
\qquad
\left\langle \Delta u\right\rangle =\Delta u+\mathcal{O}\left(  h^{2}\right).
\]
%
%
\end{document}